\documentclass[sigconf, screen, nonacm]{acmart}

\AtBeginDocument{%
  }

\usepackage{braket}
\usepackage{multirow}
\usepackage{algorithm}
\usepackage{algorithmic}
\usepackage{xcolor} 
\usepackage{array}
\newcommand{\etal}{\textit{et al}.}

\title{Genesis: A Compiler Framework for Hamiltonian Simulation on Hybrid CV-DV Quantum Computers}

\begin{document}

\author{Zihan Chen}
\authornote{These authors have equal contributions.}
\email{zihan.chen.cs@rutgers.edu}
\affiliation{
  \institution{Rutgers University}
  \city{Piscataway}
  \state{New Jersey}
  \country{USA}
}

\author{Jiakang Li}
\email{jiakang.li@rutgers.edu}
\authornotemark[1]
\affiliation{
  \institution{Rutgers University}
  \city{Piscataway}
  \state{New Jersey}
  \country{USA}
}

\author{Minghao Guo}
\email{minghao.guo@rutgers.edu}
\authornotemark[1]
\affiliation{
  \institution{Rutgers University}
  \city{Piscataway}
  \state{New Jersey}
  \country{USA}
}

\author{Henry Chen}
\email{hc867@scarletmail.rutgers.edu}
\affiliation{
  \institution{Rutgers University}
  \city{Piscataway}
  \state{New Jersey}
  \country{USA}
}

\author{Zirui Li}
\email{zl606@scarletmail.rutgers.edu}
\affiliation{
  \institution{Rutgers University}
  \city{Piscataway}
  \state{New Jersey}
  \country{USA}
}

\author{Joel Bierman}
\email{jhbierma@ncsu.edu}
\affiliation{
  \institution{North Carolina State University}
  \city{Raleigh}
  \state{North Carolina}
  \country{USA}
}

\author{Yipeng Huang}
\email{yipeng.huang@rutgers.edu}
\affiliation{
  \institution{Rutgers University}
  \city{Piscataway}
  \state{New Jersey}
  \country{USA}
}

\author{Huiyang Zhou}
\email{hzhou@ncsu.edu}
\affiliation{
  \institution{North Carolina State University}
  \city{Raleigh}
  \state{North Carolina}
  \country{USA}
}

\author{Yuan Liu}
\email{q_yuanliu@ncsu.edu}
\affiliation{
  \institution{North Carolina State University}
  \city{Raleigh}
  \state{North Carolina}
  \country{USA}
}

\author{Eddy Z. Zhang}
\email{eddy.zhengzhang@gmail.com}
\affiliation{
  \institution{Rutgers University}
  \city{Piscataway}
  \state{New Jersey}
  \country{USA}
}

\begin{abstract}
This paper introduces Genesis, the first compiler designed to support Hamiltonian Simulation on hybrid continuous-variable (CV) and discrete-variable (DV) quantum computing systems. 
Genesis is a two-level compilation system. At the first level, it decomposes an input Hamiltonian into basis gates using the native instruction set of the target hybrid CV-DV quantum computer. At the second level, it tackles the mapping and routing of qumodes/qubits to implement long-range interactions for the gates decomposed from the first level. Rather than a typical implementation that relies on SWAP primitives similar to qubit-based (or DV-only) systems, we propose an integrated design of connectivity-aware gate synthesis and beamsplitter SWAP insertion tailored for hybrid CV-DV systems.  We also introduce an
OpenQASM-like domain-specific language (DSL) named
CVDV-QASM to represent Hamiltonian in terms of Pauli-exponentials and basic gate sequences from the
hybrid CV-DV gate set.
Genesis has successfully compiled several important Hamiltonians, including the Bose-Hubbard model, $\mathbb{Z}_2-$Higgs model, Hubbard-Holstein model, Heisenberg model and Electron-vibration coupling Hamiltonians, which are critical in domains like quantum field theory, condensed matter physics, and quantum chemistry. Our implementation is available at Genesis-CVDV-Compiler \href{https://github.com/ruadapt/Genesis-CVDV-Compiler}{https://github.com/ruadapt/Genesis-CVDV-Compiler}.
\end{abstract}


\maketitle 

\section{Introduction}

To date, most quantum computing architectures are homogeneous, featuring two-state, discrete-variable (DV) realizations as qubits. The hybrid continuous-variable discrete-variable (CV-DV) quantum architecture is an emerging platform incorporating both qubits and qumodes. A qumode has a countable infinity of states in principle, thereby providing a larger Hilbert space for computation, and often has a longer lifetime than that of a qubit. As a result, qumodes have been an attractive target for quantum error correction. For instance, superconducting cavity architecture was the first to achieve memory quantum error correction above the break-even point \cite{ofek2016extending, campagne2020quantum,sivak+:nature23,ma2020error,ni2023beating,gertler2021protecting,de2022error,fluhmann2020direct} with bosonic codes. 

Most notably, a hybrid CV-DV system can simulate mixtures of fermionic and bosonic matter. Simulating physical systems has long been considered a key killer application of quantum computers, as originally proposed by Feynman \cite{feynman:ijtp82}. While DV qubits can potentially address Fermion simulation challenges, they are poorly suited for bosonic fields due to the difficulty of mapping bosonic systems to qubits and implementing bosonic field operators on infinite-dimensional Hilbert spaces. In contrast, bosonic field operators are natively available in hardware with bosonic modes or qumodes. Recent work has demonstrated the great potential of simulating Fermion-Boson mixtures on hybrid CV-DV quantum processors for applications such as material discovery \cite{knorzer2022spin}, molecular simulation \cite{wang2020efficient,wang2023observation}, topological models \cite{petrescu2018fluxon}, and lattice gauge theory \cite{crane2024hybrid}.

\begin{figure}[htbp]
\centering
\includegraphics[width=0.48\textwidth]{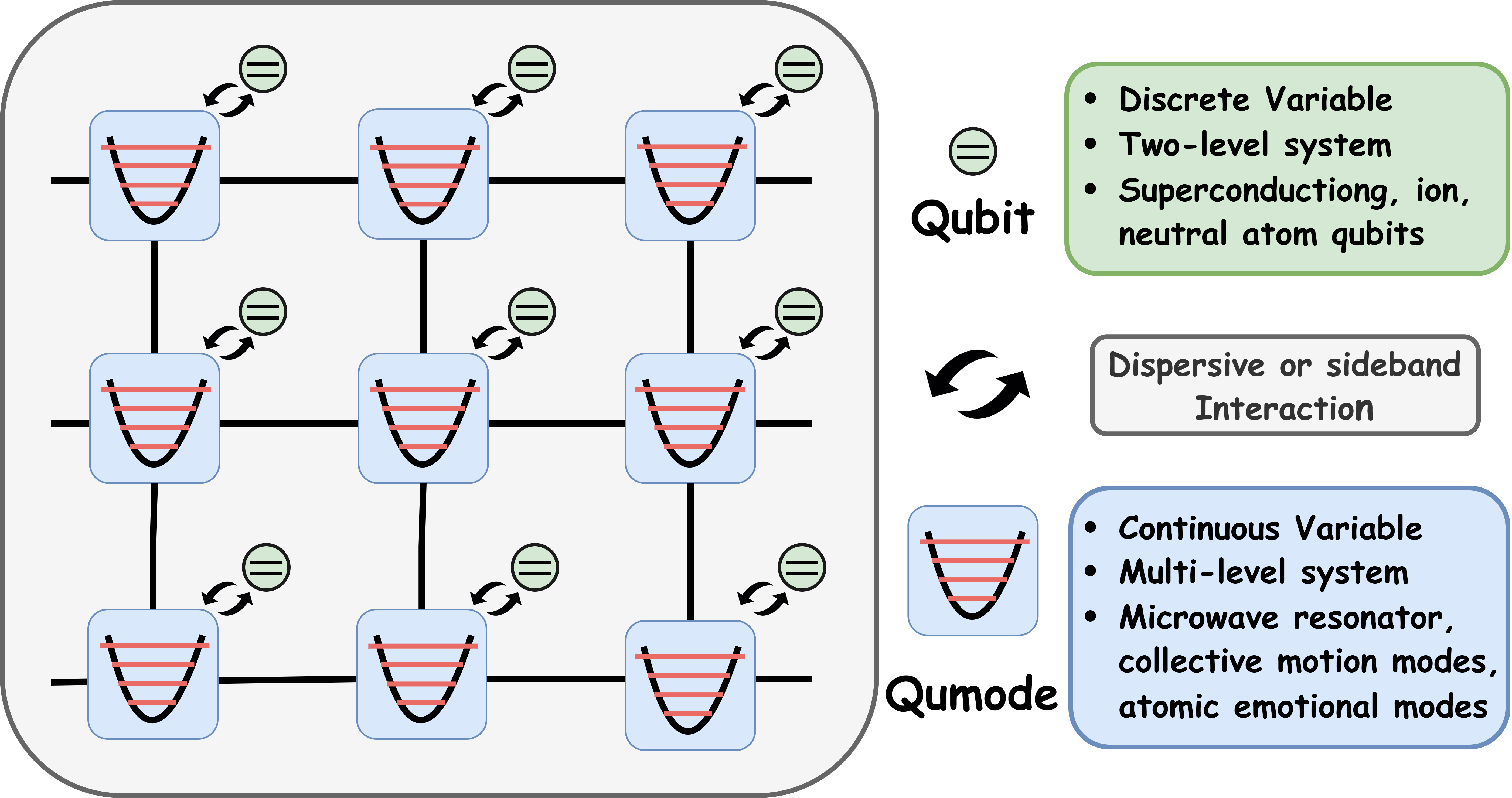}
\caption{A typical hybrid CV-DV architecture using the superconducting technology. The qumodes are connected in a sparse manner. Each qubit connects to a qumode, and there is no direct connection between qubits \cite{liu2024hybridoscillatorqubitquantumprocessors}.   }
\Description{A typical hybrid CV-DV architecture using the superconducting technology. The qumodes are connected in a sparse manner. Each qubit connects to a qumode, and there is no direct connection between qubits \cite{liu2024hybridoscillatorqubitquantumprocessors}. }
\label{fig:hybridcvdvarch}
\vspace*{-0.5\baselineskip}
\end{figure}

In this paper, we provide compiler support for mapping Hamiltonian simulation instances onto hybrid CV-DV processors. Compilation support for hybrid CV-DV architectures is still in its infancy. Existing tools, such as Bosonic Qiskit \cite{stavenger+:hpec22}, StrawberryField \cite{killoran2019strawberryfields}, Perceval \cite{zhou2024bosehedral}, and Bosehedral \cite{maring2023generalpurposesinglephotonbasedquantumcomputing}, offer preliminary support for programming, simulating, and composing circuits—primarily for domain-specific applications like Gaussian Boson Sampling (GBS) or general bosonic circuits. However, none support hybrid CV-DV Hamiltonian simulation. This compilation process requires synthesizing a circuit from the Hamiltonian operator's mathematical representation and mapping the synthesized logical circuit onto a physical circuit while ensuring compatibility with hardware constraints (an example of a superconducting hybrid CV-DV architecture is shown in Fig.~\ref{fig:hybridcvdvarch}).

Our work fills this gap. We introduce \textbf{Genesis}, the first comprehensive compilation framework for Hamiltonian simulation on hybrid CV-DV computers. It consists of two levels of compilation: \textbf{Level-1} decomposes an n-qubit-m-qumode Hamiltonian into universal basis gates, while \textbf{Level-2} compilation deals with hardware constraints, including topology constraints, multi-qubit gates, and ancilla qubit/qumode allocation. Our framework successfully compiles important Hamiltonians, such as the Bose-Hubbard model, $\mathbb{Z}_2$-Higgs model, Hubbard-Holstein model, and vibration-electron (vibronic) coupling Hamiltonians, crucial in quantum field theory, condensed matter physics, and quantum chemistry. We make the following contributions. 

\begin{figure*}[htbp]
    \centering
    \includegraphics[width=0.85\textwidth]{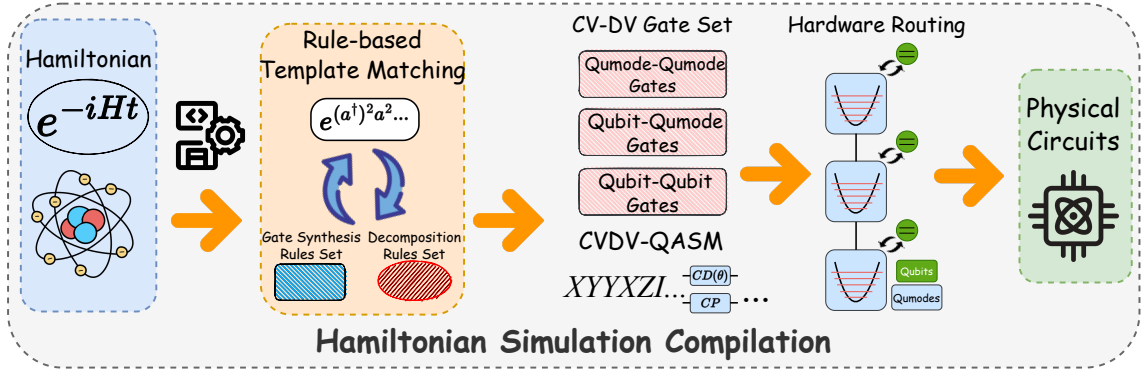}
    \caption{Compilation workflow of Genesis. It first decomposes the Hamiltonian into CV-DV basis gate sets, as well as the Pauli gate we defined in this paper, in the CVDV-QASM language format. Finally, it considers connectivity constraints, performs the hardware mapping and routing stage, and outputs physical circuits.}
    \Description{
    This figure shows the compilation workflow of Genesis, a compiler for Hamiltonian simulation on hybrid continuous-variable and discrete-variable (CV-DV) quantum systems. The process begins with a Hamiltonian expressed as \( e^{-iHt} \). In the first stage, the compiler applies rule-based template matching, using a combination of Gate Synthesis Rules and Decomposition Rules to transform the Hamiltonian into sequences of basic gates, including CV-DV gate types such as Qumode-Qumode, Qubit-Qumode, and Qubit-Qubit interactions. These are expressed in a domain-specific language called CVDV-QASM, where circuits are represented by Pauli-exponential operators and primitive gate sequences like CD and CP. The second compilation stage handles hardware routing, mapping logical qubits and qumodes to physical resources while taking connectivity constraints into account. The output is a fully compiled physical circuit ready for execution on hybrid CV-DV hardware. Genesis supports Hamiltonians from various domains, including quantum field theory and quantum chemistry.
    }
    \label{fig:flowChart}
\end{figure*}

\begin{itemize}
    \item We propose the first compiler to synthesize circuits from hybrid CV-DV Hamiltonians using a template-rewriting approach based on product formulas and trotterization. This formulation enables automatic rule search and allows further rule expansion without modifying the core algorithm.    
    \item When synthesizing gates from a Hamiltonian, the universality of bosonic systems often requires hybrid qubit-qumode operations. Therefore, the synthesis often requires the allocation of ancilla qubits or qumodes, as well as the mapping and routing of these ancilla qubits/qumodes in the computation. This feature is implemented in our compiler.  
    \item We also provide the first compilation support for multi-qubit Pauli gate implementation in a hybrid CV-DV system. As popular hybrid CV-DV architectures do not have connectivity among qubits, rather, only connectivity exists among qumodes and between qubits and qumodes, we propose to synthesize multi-qubit Pauli gates by leveraging an effect similar to phase kickback in DV systems. 
    \item Besides the compiler support, we propose a domain-specific language (DSL) design for hybrid CV-DV Hamiltonian simulation, which represents Hamiltonians as Pauli strings and basic gate sequences. 
    \item We conducted extensive experiments, evaluating benchmarks with 600-1900 multi-qubit Pauli-string Hamiltonian terms and six key Fermion-Boson Hamiltonian models. We assessed different mapping, routing, and qumode allocation strategies, including the Floating Qubit approach.
\end{itemize}

\section{Background and Motivation}
 
\subsection{Advantage of Hybrid CV-DV Systems over CV-only and DV-only Systems}
\label{sec:cvdvcomp}

Compared with CV-only or DV-only systems, hybrid CV-DV systems offer several key advantages as summarized in Table \ref{tab:comparisonSystems}. 
    

\begin{table}[h]
    \centering
    \renewcommand{\arraystretch}{1.5}
    \resizebox{0.47\textwidth}{!}{
\begin{tabular}{|p{3.4cm}|>{\centering\arraybackslash}p{2.25cm}|>{\centering\arraybackslash}p{2.1cm}|>{\centering\arraybackslash}p{2.1cm}|}
        \hline
        {\textbf{System Characteristics}} & {\textbf{Hybrid CV-DV}} & {\textbf{CV-Only}} & {\textbf{DV-Only}} \\
        \hline
        {Energy Truncation for Simulating Bosonic States \& Operators} & {Not-required} & {Not-required} & {\textbf{Required}} \\
        \hline
        {Support for Simulating Native Bosonic Operators (e.g. Square-root Factors under Fock basis)} & {Easy} & {Easy} & {\textbf{Non-trivial \cite{crane2024hybrid}}} \\
        \hline
        {Gaussian Resource Generation Difficulty} & {Easy} & {Easy} & {Easy \cite{kitaev2008wavefunction}} \\
        \hline
        {Non-Gaussian Resource Generation Difficulty} & {Easy} & {\textbf{Difficult} \cite{PRXQuantum.2.030204}} & {N/A} \\
        \hline
        {Error Channel Complexity} & {Medium} & {Low \cite{liu2024hybridoscillatorqubitquantumprocessors}} & {\textbf{{High}}} \\
        \hline
    \end{tabular}
    }
    \caption{{Hybrid CV-DV, CV-only and DV-only Systems.}}
    \label{tab:comparisonSystems}
\end{table}

    (1) \underline{Advantage in simulation:} Quantum simulation is one of the most promising applications of quantum computers. However, simulating fermion-boson mixtures is challenging for DV systems. 
    While mapping fermion operations to qubit operations is possible, representing a bosonic mode with qubits requires truncating its infinite-dimensional Hilbert space. 
    Moreover, implementing native bosonic operations in DV hardware is further complicated, for example, by the quantum arithmetic needed to realize square-root factors under the Fock basis \cite{fluhmann2019encoding, de2022error, fluhmann2020direct, crane2024hybrid}. In contrast, CV hardware employs oscillators with infinite-dimensional Hilbert space and has intrinsic support for native bosonic operators, but has limited support for modeling fermions. A hybrid CV-DV architecture combines the strengths of both: it harnesses the larger Hilbert space of the CV system while leveraging discrete qubits to perform fermion-mapped operations in the simulation of fermion-boson mixtures. 
    
    (2) \underline{Advantage in providing non-Gaussian resources:} Achieving universal CV-based quantum computation requires non-Gaussian operations, such as cubic interactions \cite{PRXQuantum.2.030204,liu2024hybridoscillatorqubitquantumprocessors}. 
    However, non-Gaussian gates are challenging to realize on CV-only platforms. Alternatively, universal control of oscillators can also be achieved by the addition of qubit-controlled oscillator gates, which are more straightforward and much less costly and have been demonstrated successfully in the lab for superconducting-cavity circuits \cite{eickbusch+:naturephysics22, teoh+:pnas23}, trapped-ion, and neutral-atom architectures \cite{bruzewicz+:apr19, fluhmann+:nature19, deneeve+:2020errorcorrectionlogicalgrid, shaw2024erasurecoolingcontrolhyperentanglementmotion, haljan+:prl05}.  

    (3)     \underline{Advantage in error modeling and QEC:} In many CV implementations (such as superconducting resonators or optical modes), photon loss (plus minor phase noise) is the primary error source, yielding Gaussian noise channels~\cite{wang2020efficient, PRXQuantum.3.020336, PhysRevLett.131.150601}. Notably, error mechanisms in CV systems are simpler for a single oscillator with many levels than for multiple qubits of equivalent total dimensionality. Using multiple qubits to represent one oscillator introduces complex error sources, such as crosstalk and correlated errors across control lines, gates, and measurements. In contrast, CV's core error model is comparatively easier to characterize and correct. This simplicity is why bosonic error-correcting codes were the first to achieve the memory break-even point \cite{sivak+:nature23, ni2023beating}. Hybrid CV-DV systems must address both oscillator photon loss and qubit errors but typically require only one (or a few) qubits per oscillator, reducing qubit crosstalk and simplifying error correction and fault tolerance.

\subsection{Hamiltonian Simulation}

Hamiltonian Simulation plays a crucial role in understanding the quantum dynamics of various systems and scenarios in quantum physics, chemistry, and materials science \cite{dong2022quantum}. A universal quantum simulator is built upon the time evolution operator \( e^{-iHt} \), where \( H \), known as \textbf{Hamiltonian}, typically represents a Hermitian operator, and \( t \) represents time (here and the rest of the paper, we take the Planck's constant as $\hbar = 1)$. The goal of the Hamiltonian simulation is to decompose the operator \( e^{-iHt} \) into a sequence of basis gates for a given hardware backend. This decomposition enables quantum computers to approximate the system's evolution over time. 

A generic n-qubit-m-qumode Hamiltonian for a hybrid CV-DV quantum system can be represented as follows: 
\begin{align}
 H=\sum^{4^n-1}_{k=0}{P_kh_k(\hat{a}_1, \hat{a}^{\dagger}_1,\hat{a}_2, \hat{a}^{\dagger}_2,\dotsi , \hat{a}_m, \hat{a}^{\dagger}_m)}
\label{eq:HamiltonianRepresentation}
\end{align}

The operator \( P_k \) represents an element of the Pauli basis ($X, Y, Z, I$) on \( n \) Qubits, where \( k = 0, 1, \ldots, 4^n - 1 \). The function \( h_k(\hat{a}_i, \hat{a}^{\dagger}_i) \) is a finite-degree polynomial in terms of the annihilation \( \hat{a}_i \) and creation operator \( \hat{a}^{\dagger}_i \) on qumode $i$. 

An example is the following spin-Holstein model on N qubits and N qumodes \cite{knorzer2022spin}. The first term acts on the $i-$th qubit and qumode jointly, and the second term acts on the $i-$th qumode:
\begin{align}
 {H} = \sum_i^N \frac{g_i}{2} {Z}_i(\hat{a}_i^\dagger + \hat{a}_i) + \sum_i^N \frac{g_i}{2} I (\hat{a}_i^\dagger + \hat{a}_i)
 \label{eq:hsexample1}
\end{align}

The compilation goal is to map the time evolution of a Hamiltonian, i.e., \( e^{-iHt} \), into a minimum and efficient set of basis gates on the hardware. The basis gates include single-qubit gate rotation, single-qumode gates, multi-qumode, and hybrid qubit-qumode gates. A representative gate set is shown in Table \ref{tab:gateSetGrouped}. Note that two-qubit gates are not included in the basis gates, as there is no direct connection between qubits.

\subsection{The Challenges for Compiling A Hybrid CV-DV Hamiltonian Application}

Compiling Hamiltonian simulation on a CV-DV system is more challenging than compiling that on a DV system. It must account for (a) unique qumode gate decomposition rules, (b) multi-qubit Pauli-string synthesis, and (3) architecture constraints when performing Qubit/Qumode mapping and routing. Specifically, we describe three fundamental differences and challenges: 

\vspace{2pt} 

\noindent \textbf{Challenge 1: \textit{Qumode-focused Gate Synthesis}}

We need to synthesize qumode-only gates for the Hermitian polynomial of annihilation and creation operators of qumodes. For simple Hamiltonian terms, we can perform pattern matching to identify the right basis gate to implement them. For instance, for the Hamiltonian in Equation \ref{eq:hsexample1}, through Trotterization~\cite{lloyd_universal_1996}, we can convert $e^{-iHt}$ into a product of two terms $e^{-i\frac{g_i}{2}Z_i(\hat{a}_i^\dagger+\hat{a_i})}t$ and $e^{-i\frac{g_i}{2}(\hat{a}_i^\dagger+\hat{a_i})}t$ for each $i$.

The first term can be pattern-matched to the Control Displacement gate $e^{\sigma_z(\alpha a^{\dagger} - \alpha^* a)}$ in Table \ref{tab:gateSetGrouped}, by setting the displacement parameter $\alpha = -i\frac{g_i}{2}$. The second term can be pattern matched to the displacement gate $e^{\sigma_z(\alpha a^{\dagger} - \alpha^* a)}$ by parameterizing the displacement $\alpha=-i\frac{g_i}{2}$. 

\begin{table}[htbp]
    \renewcommand{\arraystretch}{1.5} 
    \resizebox{0.47\textwidth}{!}{
    \begin{tabular}{|l|c|l|}
        \hline
        \textbf{Type} & \textbf{Gate Name}  & \multicolumn{1}{c|}{\textbf{Definition}} \\\hline

        \multirow{2}{*}{Qubit} 
        & $x,y$ Rotation & $r_{\varphi}(\theta)=\text{exp}\left[-i\frac{\theta}{2}(\cos\varphi\sigma_x+\sin\varphi\sigma_y)\right]$ \\\cline{2-3}
        & $z$ Rotation & $r_z(\theta)=\text{exp}\left(-i\frac{\theta}{2}\sigma_z\right)$ \\\hline

        \multirow{3}{*}{Qumode} 
        & Phase-Space Rotation  & $\text{R}(\theta)=\text{exp}\left[-i\theta a^\dagger a\right]$ \\\cline{2-3}
        & Displacement &  $\text{D}(\alpha)=\text{exp}\left[\left(\alpha a^\dagger-\alpha^*a\right)\right]$ \\\cline{2-3}
        & Beam-Splitter  & $\text{BS}(\theta, \varphi)=\text{exp}\left[-i\frac{\theta}{2}\left(e^{i\varphi}a^\dagger b+e^{-i\varphi}ab^\dagger\right)\right]$ \\\hline
        \multirow{5}{*}{Hybrid}
        & Conditional Phase-Space Rotation  & $\text{CR}(\theta)=\text{exp}\left[-i\frac{\theta}{2}\sigma_za^\dagger a\right]$ \\\cline{2-3}
        & Conditional Parity  & $\text{CP}=\text{exp}\left[-i\frac{\pi}{2}\sigma_za^\dagger a\right]$ \\\cline{2-3}
        & Conditional Displacement  & $\text{CD}(\alpha)=\text{exp}\left[\sigma_z\left(\alpha a^\dagger-\alpha^*a\right)\right]$ \\\cline{2-3}
        & Conditional Beam-Splitter &  $\text{CBS}(\theta,\varphi)=\text{exp}\left[-i\frac{\theta}{2}\sigma_z\left(e^{i\varphi}a^\dagger b+e^{-i\varphi}ab^\dagger\right)\right]$ \\\cline{2-3}
        & Rabi Interaction   & $\text{RB}(\theta)=\text{exp}\left[-i\sigma_x\left(\theta a^\dagger-\theta^*a\right)\right]$ \\\cline{2-3}
      \hline
    \end{tabular} 
    }
\caption{ 
    Basis gates in the Hybrid CV-DV System, where $\sigma_i$ terms represent Pauli terms acting on Qubits, e.g., $\sigma_z$ is the Pauli-Z operator. $a$ and $a^\dagger$ are the annihilation and creation operators acting on Qumode. Between different Qubits and Qumode is the tensor product $\otimes$. We omit the $\otimes$ symbol following the convention in the physics literature \cite{kang2023leveraging}. 
    Single-qubit or single-qumode gates typically take around 20 ns \cite{eickbusch+:naturephysics22}. A two-qumode gate or hybrid qubit-qumode gate typically runs in the range of 400$-$1000 ns \cite{eickbusch+:naturephysics22, chapman2023highonoffratiobeamsplitter}.
    }
    \vspace*{-1\baselineskip}
\label{tab:gateSetGrouped}
\end{table}

While we have a comprehensive basis gate set of Qubit and Qumode operators, not every term in a Hamiltonian (after Trotterization) can be directly mapped to a basis gate or a combination of basis gates in the quantum hardware. For instance, the term $(a^{\dagger})^2 a^2$ is a Kerr non-linearity~\cite{kang2023leveraging} that performs a simulation of the Kerr Effect in optics, and it is not easy to be directly implemented in the hardware. It is, in fact, a complex non-linear term requiring a product formula and multiple steps of gate decomposition. In the past, such decomposition was done manually by physicists or theorists. This manual decomposition approach may be time-consuming and error-prone. Furthermore, the hardware vendors typically provide an overcomplete gate set for more flexible and robust selections of gate operations. This further complicates the manual decomposition process.  

\vspace{2pt} 
  
\noindent \textbf{Challenge 2: \textit{Multi-Qubit Pauli-string Synthesis}}

Since we aim to support the simulation of a generic hybrid CV-DV Hamiltonian, we need to consider qubit-only terms -- Pauli-string terms. The simulation of Pauli-string terms \cite{li+:asplos22, jin2023tetris, li+:isca21, paykin2023pcoast} in Hamiltonians has been extensively studied for DV systems.  A Pauli-string representation denotes tensor products of Pauli-matrices. Unlike qubits that are connected in a DV system, qubits are typically not directly connected in a hybrid CV-DV system, which is built upon either superconducting or trapped-ion devices \cite{liu2024hybridoscillatorqubitquantumprocessors}. Instead, qubits are connected to qumodes, and qumodes are connected (more details discussed in Challenge 3).  A previous idea that uses a sequence of (control) displacement gates to cancel out the effect on a qumode and then ``kick back'' the phase to the qubits has successfully implemented $R_{ZZ}$, CNOT, and Toffoli gates. Inspired by this idea, we develop a method to synthesize \textbf{an arbitrary multi-qubit Pauli-string} Hamiltonian by leveraging a multi-qubit Controlled Displacement gate and a trajectory of Displacement gates with respect to different eigenvectors. Moreover, this technique could be extended to create rules for qumode operations controlled by arbitrary multi-qubit Pauli-strings. 


\vspace{2pt} 

\noindent \textbf{Challenge 3: \textit{Limited Connectivity Constraints}} 

After gate synthesis from a Hamiltonian, there are three types of gates: two-qumode gates, single-qubit/qumode gates, and hybrid qubit-qumode gates. These gates are generated at the logical level. When mapped to the physical hardware layer, there may be long-range interactions. Just like DV architectures, in the hybrid CV-DV architectures, qubits and qumodes are often not all-to-all connected. Most current hybrid CV-DV systems are designed with either only qumodes connected among themselves, or qubits connected among themselves, but not both at the same time, due to crosstalk issues \cite{liu2024hybridoscillatorqubitquantumprocessors}. We focus on the architecture where qumodes are connected, and qubits can be indirectly connected via qumodes. An example of a superconducting CV-DV device is shown in Fig. \ref{fig:hybridcvdvarch}. The reason is that a qumode has an infinite energy level and can store the information of a qubit, but not vice versa. Also, for the simulation of Fermion-Boson-mixture, the connectivity among qumodes naturally allows the modeling of bosonic interactions. 

Qubit/qumode mapping and routing are tightly connected with circuit synthesis described in \textbf{Challenge 1 and 2}. First of all, the synthesis rules allow certain flexibilities, such as the commutation of gates, which can be leveraged to improve the mapping and routing stage to reduce the number of intrinsic SWAP gates. Moreover, the synthesis approach often requires ancilla qumodes, and most importantly, in certain cases, such as compiling multi-qubit entanglement operations with control displacement, it does not require the ancilla qumode to be in the vacuum (ground) state.
Rather, it can return the ancilla qumode back to its original state after the sequence is completed for synthesizing a qubit/qumode operation of interest. That means any qumode can be used to assist synthesis and also can be leveraged to improve the mapping and routing efficiency in the hardware circuit compilation stage. This is different from mapping and routing in the traditional DV system, where only SWAP insertion is considered. In certain cases, it can be modeled as a traveling salesman problem (TSP) as described in our Section \ref{sec:TSPformulation}. We develop a synergistic mapping and routing approach that considers the flexibilities in circuit synthesis. 

\textit{Summary:} Overall, our work is \textbf{the first comprehensive compiler framework that addresses the compilation of Hamiltonian simulation on hybrid CV-DV architectures}.  It addresses both the gate synthesis and hardware mapping problems, as well as identifies and models the unique compilation problems that only arise in the hybrid CV-DV architecture rather than in the traditional DV context. We show our workflow in Fig. \ref{fig:flowChart}, which shows the process of how a given Hamiltonian is decomposed into physical circuits step by step by our methods. We present our design details in Section \ref{sec:compframe} corresponding to each of the three challenges.

\subsection{Current State of CV-DV Technologies}
\label{sec:cvdvstate}
CV-DV systems can be implemented with superconducting, trapped-ion, or neutral atom devices. In superconducting devices, transmons are used as two-level systems (DV), and microwave resonators storing photons are used as harmonic oscillators (CV). 
In trapped-ion devices, collective motional modes are oscillators, while ions’ internal degrees of freedom are used as qubits. 
With neutral atoms, atomic motional modes in the optical tweezer are oscillators, and neutral atoms’ internal degrees of freedom are used as qubits. 

All three technologies have been demonstrated successfully in labs. Superconducting devices typically operate in GHz, trapped-ion in MHz, and neutral atom in KHz. 
Among them, superconducting devices offer advantages compared to the other two architectures, such as fast and reliable basis gate sets. 
For instance, microwave-controlled beam-splitter gates permit fast, high-fidelity oscillator SWAPs, usually in ~100 ns, with 99.92\% fidelity in the single photon subspace~\cite{lu2023high}. 
Hence, the superconducting architecture is our primary evaluation platform in this work, and our work can be extended to other hybrid CV-DV architectures. 

\section{Our Compilation Framework}
\label{sec:compframe}

As shown in Equation~\ref{eq:HamiltonianRepresentation}, the Hamiltonian consists of contributions from Pauli operators ($X, Y, Z, I$) and qumode operators ($a_i^{\dagger}, a_i$). We synthesize them into a physical circuit with basis gates while considering hardware connectivity constraints in CV-DV systems.

\begin{table*}[htbp]
    \centering
    \renewcommand{\arraystretch}{1.5}
    \resizebox{1.0\textwidth}{!}{
    
    \begin{tabular}{|c|c|c|c|c|c|}
        \hline
        \textbf{Rules} & \textbf{Operator Template} & \textbf{Conditions} & \textbf{Decomposition Output} & \textbf{Reference} & {\textbf{Precision}} \\
        \hline
        1 & $\text{exp}(Mt + Nt) \approx \text{Trotter}(Mt, Nt)$  &  & ${(\text{exp}(Mt/k)\text{exp}(Nt/k))}^k$ & Trotterization & {Approx} \\
        \hline
        2 & $\text{exp}([Mt, Nt]) \approx \text{BCH}(Mt, Nt)$ &  &  $ \text{exp}(Mt)\text{exp}(Nt)\text{exp}(-Mt)\text{exp}(-Nt)$ & BCH & {Approx} \\
        \hline
        3 & $ \text{exp}(t^2[M,N])$ & $M, N$ Hermitian & $\text{exp}([it\sigma_i N, it\sigma_i M])$ & \cite{kang2023leveraging} & {Exact} \\
        \hline
        4 & $ \text{exp}(-it^2\sigma_i\{M,N\} )$ & $M, N$ Hermitian & $\text{exp}([it\sigma_j M, it\sigma_k N])$ & \cite{kang2023leveraging} & {Exact} \\
        \hline
        5 & $\text{exp}(-i t^2 \sigma_z[M, N])$ &  & $\text{exp}([(it N, it\sigma_z M])$ & This paper & {Exact} \\
        \hline

        6 & $\text{exp}(t^2 \sigma_z((MN-(MN)^\dagger)))$ & $[M,N]=0$ & $\text{exp}([X \cdot it\mathcal{B}_N\cdot X,it\mathcal{B}_M]$) & \cite{kang2023leveraging} & {Exact} \\
        \hline
        7 & $\text{exp}(it^2 \sigma_z((MN+(MN)^\dagger)))$ & $[M,N]=0$ & $\text{exp}([S\cdot it\mathcal{B}_M \cdot S^{\dagger} ,X \cdot it\mathcal{B}_N\cdot X])$ & \cite{kang2023leveraging} & {Exact} \\
        \hline
        8 & $\text{exp}\left(-2it \begin{pmatrix}
            MN & 0 \\
            0 & -MN
        \end{pmatrix}\right)$ & $M, N$ Hermitian & $\text{exp}(-it\sigma_z[M,N] -it \sigma_z\{M, N\})$  & This paper & {Exact} \\
        \hline
        9 & $\text{exp}\left(2it^2 \begin{pmatrix}
            MN & 0 \\
            0 & -MN
        \end{pmatrix}\right)$ & 
        \(\begin{aligned}
            &[M,N]=0 \\
            &MN=(MN)^\dagger
        \end{aligned}\) & $\text{exp}([(S\cdot it \mathcal{B}_M \cdot S^{\dagger}, X\cdot it \mathcal{B}_N \cdot X])$& \cite{kang2023leveraging} & {Exact} \\
        \hline
        10 & $\text{exp}\left(2it \mathcal{B}_{MN} \right)$ & $[M, N]=0$ & $X\cdot \text{exp}(t\sigma_y(MN-(MN)^\dagger) + it\sigma_x(MN+(MN)^\dagger) ) \cdot X$  & \cite{kang2023leveraging} & {Exact} \\
        \hline
        11 & $\text{exp}\left(it \begin{pmatrix}
            2MN & 0 \\
            0 & -NM-(NM)^{\dagger}
        \end{pmatrix}\right)$ & $MN = (MN)^{\dagger}$ & $\text{exp}([S\cdot it \mathcal{B}_M \cdot S^{\dagger}, X\cdot it \mathcal{B}_N \cdot X])$ & \cite{kang2023leveraging} & {Exact} \\
        \hline
        12 & $\mathcal{B}_{a} = \text{exp}\left(2i\alpha \begin{pmatrix}
            0 & a \\
            a^{\dagger} & 0
        \end{pmatrix}\right)$ & $\alpha=\alpha^*$ & $ \text{exp}(i(\pi/2)a^\dagger a)\text{exp}(i(\alpha(a^\dagger+a))\otimes \sigma_y)\text{exp}(-i(\pi/2)a^\dagger a)\text{exp}(i(\alpha(a^\dagger+a))\otimes \sigma_x) $ & \cite{kang2023leveraging} & {Approx} \\
        \hline
        13 & $\mathcal{B}_{a^{\dagger}} = \text{exp}\left(2i\alpha \begin{pmatrix}
            0 & a^{\dagger} \\
            a & 0
        \end{pmatrix}\right)$ & $\alpha=\alpha^*$ & $ \text{exp}(i(\pi/2)a^\dagger a)\text{exp}(i(\alpha(a^\dagger+a))\otimes \sigma_y)\text{exp}(-i(\pi/2)a^\dagger a)\text{exp}(-i(\alpha(a^\dagger+a))\otimes \sigma_x) $ & This paper & {Approx} \\
        \hline
        {14} & {$e^{(P_1P_2\cdots P_n)(\alpha a_k^\dagger - \alpha^* a_k)}$} & {} & {Multi-qubit-controlled displacement: Right hand side (RHS) of Equation (\ref{eq:multiqubitCD}) first line} & {\cite{liu2024hybridoscillatorqubitquantumprocessors}} & {Exact} \\ \hline
        {15} & {$e^{2i\alpha^2 P_1P_2\cdots P_n}$} & {} & {Multi-Pauli Exponential: Right hand side (RHS) of Equation (\ref{eq:entireseq}) first line} & {This Paper} & {Exact} \\ \hline
        {16} & {All Native Gates RHS in Table \ref{tab:gateSetGrouped}} & {} & {All Native Gates Left Hand Side (LHS) Table \ref{tab:gateSetGrouped}} & {\cite{liu2024hybridoscillatorqubitquantumprocessors}} & {Exact} \\ \hline
    \end{tabular} 
    }
    \caption{Decomposition rules for operators in hybrid CV-DV systems with corresponding conditions. \( X \) and \( S \) represent Pauli-X and phase gates, implemented using the single-qubit rotation gates in Table~\ref{tab:gateSetGrouped}.  All rules are exact except rules 1, 2, 12, and 13, which are subject to Trotter and BCH approximation errors with respect to the order of Trotter or BCH. 
    }
    \vspace*{-0.5\baselineskip}
    \label{tab:operatorDecompositionRules}
\end{table*}

\subsection{Qumode Gate Synthesis}
\label{sec:autoSynthesis}

\subsubsection{Pattern Matching}
\label{sec:gateDecomposition}

For sequences involving first or second-order Hermitian polynomials of annihilation/creation operators, we first try to map them to the basic gates in the hybrid CV-DV gate set (Table~\ref{tab:gateSetGrouped}) through pattern matching. The corresponding basic gates are applied directly if the operators align with specific gate patterns with parameterization -- most of the bosonic gates have continuous parameters. These sequences can then be synthesized into implementable gate operations. For example, common terms like \( a^{\dagger}a \) and \( (a^{\dagger} - a) \) can be mapped to the Phase-Space Rotation gate and the Displacement gate, respectively, as shown below:
\begin{equation}
    \begin{split}
        e^{(-ti a^{\dagger}a)} &\rightarrow R(t),~\text{where}~R(\theta) = e^{-i\theta a^{\dagger}a} \\
    e^{(3ia^{\dagger}+3ia)} &\rightarrow D(3i),~\text{where}~ D(\alpha) = e^{\alpha a^{\dagger} - \alpha^* a}
    \end{split}
\end{equation}
The basic gate set in Table~\ref{tab:gateSetGrouped} defines the matching rules, which allow further extension for wider functionally or architecture-specific gates. However, it only works for simple and fine-grained qumode operator sequences. For complex and large-scale annihilation/creation operator polynomials, we propose the following recursive decomposition process to synthesize them.

\subsubsection{Template Matching Decomposition and Approximation}
\label{sec:templateMatching}

Most Hamiltonians are not directly implementable on quantum hardware and require further decomposition and approximation. Two common methods for Hamiltonian decomposition are Trotterization and the Baker Campbell Hausdorff (BCH) formula, both widely used in quantum circuit synthesis \cite{kang2023leveraging,li2024principal, aftab2024multi, mansuroglu2023problem}.

\textbf{Trotterization:} The Trotter-Suzuki decomposition approximates the time evolution operator by breaking it into simplified exponentials:
\begin{align}
    e^{(M+N)t} \approx \left(e^{Mt'} e^{Nt'}\right)^n,
\end{align}
where $M$ and $N$ are parts of the Hamiltonian, $t' = t/n$, and $n$ is the number of Trotter steps. This method is particularly effective for approximating long-time evolution by transforming it into discrete, simplified steps. The purpose of Trotterization is to break a Hamiltonian into a \textbf{sum} of terms, which later can be used for approximation as products of small matrix exponentials.

\textbf{Baker Campbell Hausdorff (BCH):} The BCH formula is used to decompose the time evolution operator with non-commuting terms, allowing for a systematic approximation:
\begin{align}
    e^{[M,N]t^2} \approx e^{Mt} e^{Nt} e^{-Mt} e^{-Nt}.
\end{align}
where $[M, N] = MN - NM$ is the commutator. This approach is most suitable for short-time evolution or small non-commutative contributions. BCH is useful, as it can be used to help create a \textbf{product} of terms.

    Both Trotterization and BCH can be approximated to an arbitrary precision with a large enough order. 
    For the BCH formula, we use the second-order approximation. Trotter will affect the number of gates by repeating the same gate sequence $n$ times, but each gate has a smaller parameter and has a smaller duration. The overall evolution time is the same after Trotterization. 

Consider the following commutator (assuming A and B are qumode operators):
\begin{align}
    [\sigma_z \otimes A, I \otimes B] = \sigma_z \otimes AB - \sigma_z \otimes BA = \sigma_z [A, B]. 
    \label{eq:comm}
\end{align}

Now also consider that $([A,B] + \{A, B\})/2 = AB$, where $\{A, B\} = AB + BA$ is the anticommutator, the implication is that \textbf{we can implement a product of terms of $\hat{a}$ and $\hat{a}^\dagger$ operators} if $\sigma_z\{A, B\}$ can be implemented. Fortunately, Kang \etal~ \cite{kang2023leveraging} provides an implementation of $\sigma_z\{A, B\}$ on CV-DV architecture, as well as another version of $[A, B]$ implementation, which is different from ours in Equation \ref{eq:comm}. These altogether set a foundation for the template-rewriting-based approach we propose in this paper. We list all these rules in Table \ref{tab:operatorDecompositionRules}. With basic \textbf{sums} and \textbf{products} rules, we can recursively break a Hamiltonian down into sums and products until we reach a point where the term of interest matches a basis gate template in Table \ref{tab:gateSetGrouped}. We describe the recursive template matching process in the next section. 

In Table~\ref{tab:operatorDecompositionRules}, For a qubit$\otimes$qumode system, the Hamiltonian pattern \( \mathcal{B}_M \) :
\[
\mathcal{B}_M = \begin{pmatrix}
0 & M \\
M^\dagger & 0
\end{pmatrix},
\]
as a Hermitian operator acting on the combined Hilbert space \( \mathcal{H}_2 \otimes \mathcal{H}_{\Lambda+1} \), where \( \mathcal{H}_2 \) is the qubit Hilbert space and $\mathcal{H}_{\Lambda+1}$ is the qumode Hilbert space.

Certain decomposition rules in Table \ref{tab:operatorDecompositionRules} represent that the target operator is in a block encoding form. For instance, we only need the $MN$ component in the top-left of the Hamiltonian below. 
 \begin{align}
 \begin{pmatrix}
    2MN & 0 \\
    0 & -NM - (NM)^\dagger
\end{pmatrix}.
\end{align}
In a qubit-qumode system, an operator \( \mathcal{O} = \begin{pmatrix} A & B \\ C & D \end{pmatrix} \) can be expressed as:
\[
\mathcal{O} = \ket{0}\bra{0} \otimes A + \ket{0}\bra{1} \otimes B + \ket{1}\bra{0} \otimes C + \ket{1}\bra{1} \otimes D.
\]

Projecting onto the subspace associated with \( \ket{0} \) by setting the ancilla qubit to \( \ket{0} \), the effective operator becomes \( \langle 0 | \mathcal{O} | 0 \rangle = A \), assuming \( C = 0 \). Terms involving \( \ket{0}\bra{1} \) and \( \ket{1}\bra{1} \) vanish due to orthogonality. Thus, by initializing the ancilla and zeroing specific components (e.g., \( C = 0 \)), we can isolate and simplify blocks of the operator matrix for easier decomposition. If we take the matrix exponential of $\mathcal{O}$, let $B=C=0$, since $A$ is on the diagonal, we can also implement $e^A$ via $e^{\mathcal{O}}$ via the block encoding. 

\subsubsection{Rule-Based Recursive Template Matching for Hamiltonian Decomposition}
\label{sec:recursive}

Section~\ref{sec:gateDecomposition} introduces template matching rules for synthesizing operators into gates, while Section~\ref{sec:templateMatching} extends these rules to decompose operator sequences. To decompose complex Hamiltonians, \textbf{recursive steps} and \textbf{multiple decomposition paths} are needed. To enable scalable and automated Hamiltonian simulation, we transform the decomposition task into a \textbf{Template Matching compilation}. This recursive process applies gate synthesis and decomposition rules iteratively to input sequences of annihilation and creation operators so as to reduce a Hamiltonian to basic gates in the Hybrid CV-DV gate set.

Next, we use an example, Hamiltonian in Equation~\ref{eq:nonlinearH}, to illustrate the decomposition process. We first apply Rule 1, the Trotterization formula, to approximate it as a product of two exponential operators. As Rule 1 and Rule 2 are common \textbf{intermediate steps} in many decomposition Rules, we omit their detailed discussion here.
\begin{equation}
\begin{split}
        \text{exp}(-iHt) &= \text{exp}\big(-i(\omega a^{\dagger}a + \frac{\kappa}{2}(a^{\dagger})^2 a^2)t\big) \\
    &\approx \left(\text{exp}(-i\omega a^{\dagger}a \frac{t}{k}) \cdot \text{exp}(-i\frac{\kappa}{2}(a^{\dagger})^2 a^2 \frac{t}{k})\right)^k
\end{split}    
\label{eq:nonlinearH}
\end{equation}
Rule 1 separates the linear term \( a^{\dagger}a \) and nonlinear term \( (a^{\dagger})^2 a^2 \). Each term is then decomposed into basic gates in the Hybrid CV-DV gate set. For the linear term \( a^{\dagger}a \), it is synthesized as a Phase-Space Rotation gate using the rules from Section~\ref{sec:gateDecomposition}.

Decomposing the nonlinear term \((a^{\dagger})^2 a^2\) is challenging due to the lack of a matching basis gate. This would incur recursively applying the template matching rules from Section~\ref{sec:templateMatching} or Table \ref{tab:operatorDecompositionRules} to simplify it. The process follows a recursive search tree, where nodes represent decomposition states (current terms and their settings), and edges correspond to synthesis rules from Tables~\ref{tab:gateSetGrouped}-\ref{tab:operatorDecompositionRules}. 

\begin{figure}[h]
    \centering
    \includegraphics[width=0.35\textwidth]{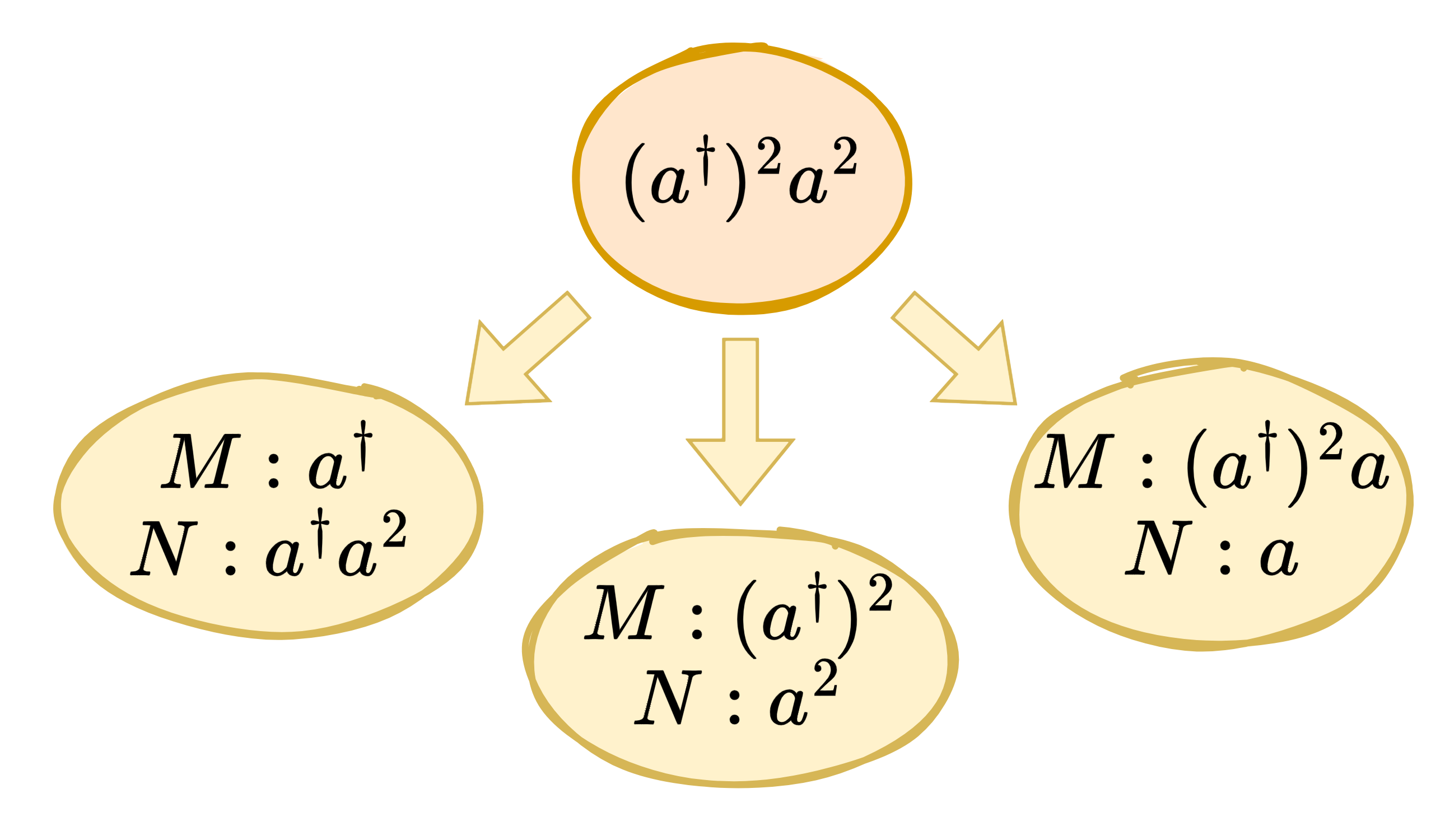}
    \caption{Decomposing $(a^{\dagger})^2 a^2$ into three child states, i.e., splitting it into subterms $M$ and $N$ using three methods.}
    \Description{Decomposing $(a^{\dagger})^2 a^2$ into three child states, i.e., splitting it into subterms $M$ and $N$ using three methods.}
    \label{fig:1to3}
\end{figure}

For a complex term, there may exist multiple ways to break it into subterms, \(M\) and \(N\) in Table~\ref{tab:operatorDecompositionRules}.  
Fig.~\ref{fig:1to3} shows three possible methods for splitting \((a^{\dagger})^2 a^2\) into \(M\) and \(N\). This introduces additional complexity or opportunity to the decomposition process due to the multipath selection.

To decompose \((a^{\dagger})^2 a^2\), we select a splitting method for \(M, N\), then apply the template matching rules from Table~\ref{tab:operatorDecompositionRules}. First, we check the conditions for each rule and filter out those that don't apply. Next, we identify the operator templates for \(M\) and \(N\), such as \(MN\), \([M,N]\), \(\{M,N\}\), \(MN \pm (MN)^{\dagger}\), or \(\mathcal{B}_{MN}\), and proceed with the decomposition accordingly.

We use Depth-First Search (DFS) with backtracking to traverse the recursive decomposition tree, exploring paths deeply and storing valid decomposition paths for evaluation. Dead ends occur when no rule further reduces the operator sequence. The algorithm then backtracks to the last valid node, prunes unnecessary branches, and explores alternative paths.

\textbf{Handling Dead Ends:} For instance, if \( M = a^\dagger \) and \( N = a^\dagger a^2 \), Rule 11 from Table~\ref{tab:operatorDecompositionRules} applies as \( MN = (MN)^\dagger \). While \( M \) matches a basic gate, \( N \) (\( a^\dagger a^2 \)) requires further decomposition. The path reaches a dead end after trying three splitting methods for \( N \) without success. The algorithm then discards this branch and backtracks to try an alternative decomposition, such as \( M = (a^\dagger)^2 \) and \( N = a^2 \).

\textbf{Step 1:}  
For \( M = (a^\dagger)^2 \) and \( N = a^2 \), the condition \( MN = (MN)^\dagger \) holds, allowing Rule 11 to be applied. This decomposes the term into two subterms, \( (S \cdot it \mathcal{B}_M \cdot S^\dagger) \) and \( (X \cdot it \mathcal{B}_N \cdot X) \). 

\textbf{Step 2:}  
One of subterm \( (a^\dagger)^2 \) is further decomposed into \( M = a^\dagger \) and \( N = a^\dagger \), satisfying the condition \( [M, N] = 0 \) and matching the Block Encoding template. Rule 10 is applied to produce two new subterms: \( it \sigma_x(MN + (MN)^\dagger) \) and \( t \sigma_y(MN - (MN)^\dagger) \).

\textbf{Step 3:}  
At this stage, \( M = a^\dagger \) and \( N = a^\dagger \) again. Rule 6 is applied, decomposing the term into \( (S \cdot it \mathcal{B}_M \cdot S^\dagger) \) and \( (X \cdot it \mathcal{B}_N \cdot X) \), further simplifying the operators.

\textbf{Step 4:}  
The Block Encoding template is then decomposed into a sequence of simplified annihilation and creation operators using Rule 13.

\begin{figure}[h]
    \centering
    \includegraphics[width=0.45\textwidth]{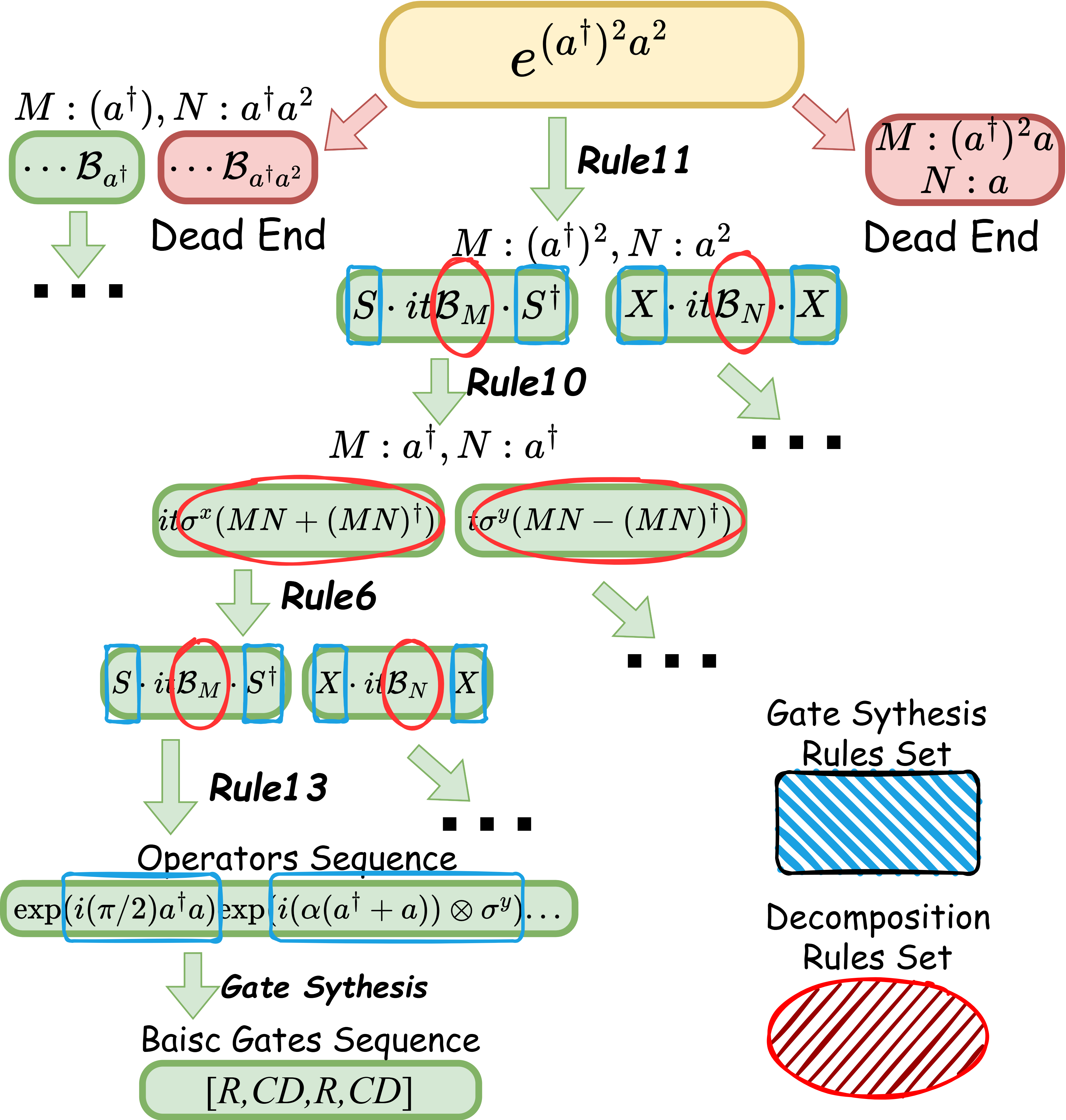}

    \caption{Recursive decomposition and gate synthesis of $(a^{\dagger})^2 a^2$ into a sequence of basic gates. 
    The red circle indicates the template matching rules from the Decomposition Rules Set (Table~\ref{tab:operatorDecompositionRules}), and the blue square represents the template matching rules from the hybrid CV-DV gate set (Table~\ref{tab:gateSetGrouped}).}
    \Description{
This figure illustrates the recursive decomposition and gate synthesis of the operator \( e^{(a^\dagger)^2 a^2} \) into a sequence of basic gates. The process explores multiple paths. The valid decomposition path applies Rule 11 to reduce the expression to terms involving \( M : (a^\dagger)^2, N : a^2 \). Rule 10 further transforms it into operator sequences such as \( \exp(i\theta(MN + (MN)^\dagger)) \) and \( \exp(i\theta(MN - (MN)^\dagger)) \). Rule 6 decomposes these into intermediate gates, including sequences like \( S, iB_M, S^\dagger, X, iB_N, X \). Finally, Rule 13 synthesizes an operator sequence \( \exp(i \frac{\pi}{2} a^\dagger a) \exp(i\theta(a^\dagger + a) \otimes \sigma^\nu) \), which is mapped into a basic gate sequence \([R, CD, R, CD]\). Red circles denote template matches from the Decomposition Rules Set (Table 3), and blue squares indicate matches from the hybrid CV-DV gate set (Table 2).
}
    \vspace*{-0.75\baselineskip}
    \label{fig:5steps}
\end{figure}

\textbf{Step 5:}  
Finally, the resulting sequence is synthesized into Phase-Space Rotation and Displacement gates from the hybrid CV-DV gate set. The complete 5-step recursive Template Matching decomposition and gate synthesis process is illustrated in Fig.~\ref{fig:5steps}.

\noindent \underline{\textbf{Discussion:}} Our mechanism is designed to be \textbf{extension-friendly}, \textbf{reuse-friendly}, and \textbf{customization-friendly}. When expanding two rule sets, such as adding new architecture-specific gates or decomposition rules, our template matching and implementation mechanism remains unaffected. Additionally, we can customize cost metrics, such as fidelity or decomposition approximation, by extending our template-matching into a cost-optimized framework. 

Hamiltonian decomposition and gate synthesis are highly challenging tasks. To the best of our knowledge, our work is the first to automatically decompose the six general Hamiltonian models described in Section \ref{sec:eval} for the CV-DV system. 
Our evaluation indicates that the decomposition and synthesis of a given Hamiltonian operator have limited viable pathways. As a result, the trade-offs between latency, gate count, and error rate are not a primary concern at this stage, given the scarcity of successful decomposition pathways. However, providing support to choose among different decomposition and synthesis pathways should be a feasible extension. We leave this extension of our compiler as our future work. 

\subsection{Multi-qubit Pauli-string Synthesis}
\label{sec:multi-qubit-gate-synthesis}

We propose a scheme to synthesize an arbitrary multi-qubit Pauli-string on hybrid CV-DV platforms. It is inspired by \textit{phase kickback} in DV systems, where the phase of the control qubit is influenced by the operation on the target qubits.  In our approach, we use qumodes as a medium to pass entanglement among qubits through a sequence of Conditional Displacement gates, i.e., qubit-controlled qumode Displacement, and unconditional Displacement gates such that the desired operation is achieved on qubits while the effects on qumodes cancel.

We propose, for the first time, the multi-Pauli exponential for the CV-DV system in this paper, that is, an arbitrary Pauli-string exponential for a hybrid CV-DV system built upon displacement and multi-qubit control displacement gates. 
 
The detailed implementation follows a structured decomposition:
\begin{equation}
    \begin{split}
    U &= D^{k}(i\alpha) \, CD^{(k, P_{1\cdots n})}(-\alpha) \, D^{k}(-i\alpha) \, CD^{(k, P_{1\cdots n})}(\alpha) \\
    &= e^{2i\alpha^2 P_1P_2\cdots P_n}
    \end{split}
    \label{eq:entireseq}
\end{equation}
where $P_1P_2\cdots P_n$ represents the desired multi-qubit Pauli string \( P_1 \otimes P_2 \otimes \cdots \otimes P_n \), \( D^{k}(m) \) is a displacement of \( m \) to the \( k \)-th qumode, and $CD^{(k, P_{1\cdots n})}(-\alpha)$ and $CD^{(k, P_{1\cdots n})}(\alpha)$ are controlled displacement of $\pm\alpha$. During the operation, displacement effects on qumodes return them to the original states, kicking back an overall phase to the qubits, akin to phase kickback in DV systems. This is built on Liu \etal \cite{liu2024hybridoscillatorqubitquantumprocessors}, which used ancillary qumodes to implement CNOT and $R_{ZZ}$ and single-qubit gates. We extend their approach to cover arbitrary multi-qubit operations with machine compilation. 

An example with one qubit and one qumode is as follows. 
\begin{equation}
    \begin{split}
    U&=D^{k}(i\alpha)CD^{k, P_1}(\alpha,-\alpha)D(-i\alpha)CD^{k, Z}(-\alpha,\alpha) \\ &= e^{2i\alpha^2Z}
    \end{split}
    \label{eq:multiqubitgate}
\end{equation}
If we set $P_1$ as the Pauli-Z operator, \( CD^{k, Z}(m) \) applies a qubit-controlled Displacement gate to the \( k \)-th qumode, with \( D(m) \) for \( \ket{0} \) and \( D(-m) \) for \( \ket{1} \) with respect to the state of the first qubit, this exactly implements $e^{2i\alpha^2 Z}$. 

Our new multi-Pauli exponential in Equation (\ref{eq:entireseq}) makes use of a multi-qubit controlled CD gate proposed by Liu \etal  \cite{liu2024hybridoscillatorqubitquantumprocessors}, as below. 

\begin{equation}
\begin{split}
\text{CD}^{(k, P_1 P_2 \cdots P_n)} (\alpha) &= U_{\text{seq}}^\dagger D(i^n \alpha) U_{\text{seq}} \\
&= e^{(P_1P_2\cdots P_n)(\alpha a_k^\dagger - \alpha^* a_k)}
\end{split}
\label{eq:multiqubitCD}
\end{equation}
where \( U_{\text{seq}} = \prod_{j}^n e^{i \pi P_j a_k^\dagger a_k / 2} \) is a sequence of Control Parity gates (in Table \ref{tab:gateSetGrouped}) conjugated by Clifford gates, with the $j-$th qubit controlling the $k-$th qumode. It acts as a Control Parity gate for \( P_j = Z \). For \( P_j = X \), the Control Parity gate must be conjugated with Hadamard (\( H \)) gates, and for \( P_j = Y \), with \( H \) and \( S^\dagger \) gates.

Essentially, this method combines unconditional Displacement and Control Parity gates to construct an arbitrary multi-qubit Pauli-string term, requiring only a single ancillary qumode. The approach is \textbf{agnostic to the state of the ancilla qumode}, with the ancilla qumode returning to its initial state after the operation. This flexibility supports \textbf{efficient resource allocation} and \textbf{specific optimization goals}, benefiting mapping and routing.

\begin{figure}[htbp]
    \centering
    \includegraphics[width=0.45\textwidth]{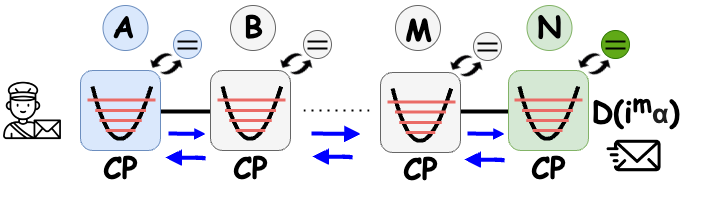}
    \caption{An ancilla qumode interacts with every qubit on the path from A to N via Control Parity (CP) gates, performs an unconditional Displacement gate, and then travels back to A's location while performing CP gates. This implements a CD gate in Equation \ref{eq:multiqubitCD}. Note that the ancilla qumode can be any qumode, and the CP gates can run in any order. We just pick the qumode below qubit A to illustrate this idea.}
    \vspace*{-0.5\baselineskip}
    \Description{
    This figure shows an ancilla qumode interacting with every qubit from A to N using Control Parity (CP) gates. The qumode begins below qubit A, travels through qubits B and M, and finally interacts with qubit N. During the forward path, the qumode performs CP gates with each qubit. At the end of the path, it undergoes an unconditional Displacement gate labeled D(i*m*α). It then returns to A’s location, again applying CP gates along the way. This bidirectional interaction sequence implements a CD gate as described in Equation 11. The qumode used can be any available one, and the CP gates can be applied in any order. The qumode shown under qubit A is chosen just for illustration.
    }
    \label{fig:multi-qubit-gate}
\end{figure}

The example in Fig.~\ref{fig:multi-qubit-gate} illustrates how an ancilla qumode travels in the physical circuit, visiting every qubit $A$, $B$, $\cdots$, $N$, to implement a CD gate for all qubits involved in the Pauli-string. 

Note that conjugating the Displacement gate with conjugated CP gates, or applying it to a polynomial of annihilation and creation operators, yields multi-qubit controlled qumode gates. These form a series of sub-rules, which are also included in our rule database (Table~\ref{tab:operatorDecompositionRules}) for conciseness.

\subsection{DSL Design and Implementation} \label{sec:DSL}

After level-1 compilation, Hamiltonians are decomposed into sequences of basic gates from the hybrid CV-DV gate set. Level-2 compilation then applies hardware constraints to generate physical circuits. To streamline this process, we introduce \textbf{CVDV-QASM}, an OpenQASM-like DSL that integrates Pauli-strings and CV-DV gate sequences. Our mapping and routing stage parses CVDV-QASM into quantum circuits. Syntax support for Pauli-string terms like ``pauli($\pi/4$) YYZI'' enables CP gate commutation and qumode allocation flexibility during physical circuit mapping while preserving semantics and providing crucial information to the lower compilation stack. An example is shown in Fig.~\ref{fig:cv-dv-dsl-example}:

\begin{figure}[htbp]
    \centering
    \begin{minipage}{0.5\textwidth}
    \setlength{\baselineskip}{0\baselineskip} 
    \begin{verbatim}
        // Pauli String with Parameter
        pauli(pi/4) YIYZXXIIIIIIII;
        pauli(pi/4) XZYZXYIIIIIIII;
        ...
        // Phase Space Rotation Gate
        R(pi/4) qm[1];
        R(pi/4) qm[2];
        ...
        // Control Displacement Gate
        CD(pi/4) q[2], qm[1];
        CD(pi/4) q[3], qm[1];
        ...
        // Displacement Gate
        D(pi/4) qm[2];
        D(pi/4) qm[2];
        ...
    \end{verbatim}
    \end{minipage}
    
    \vspace*{-0.75\baselineskip} 
    
    \caption{Hybrid CV-DV circuit DSL example. \texttt{pauli} represents the parameter of a Pauli String; \texttt{R, CD, D} represent the gate type; \texttt{qm[i]} and \texttt{q[i]} represent the qumode and qubit respectively.
    Pauli String sequence will be further decomposed into basic gates in the compiler output of physical circuits.}
    \Description{Hybrid CV-DV circuit DSL example. \texttt{pauli} represents the parameter of a Pauli String; \texttt{R, CD, D} represent the gate type; \texttt{qm[i]} and \texttt{q[i]} represent the qumode and qubit respectively.
    Pauli String sequence will be further decomposed into basic gates in the compiler output of physical circuits.}
    \label{fig:cv-dv-dsl-example}
\end{figure}

This DSL has a similar input format of Bosonic Qiskit~\cite{stavenger+:hpec22}, except that our DSL supports Pauli string representation. As described before, Pauli strings are decomposed with an ancilla qumode, while hardware constraints, such as qumode interactions, are managed by inserting SWAP gates. The final circuit after our mapping and routing consists only of basic CV-DV gates and not Pauli-string gates. Detailed discussions on hardware constraints are in Section~\ref{sec:hardwareConstraints}.

\subsection{Tackling Limited Connectivity Constraints}
\label{sec:hardwareConstraints}

In hybrid CV-DV architectures shown in Fig. 1, qumodes have limited connectivity and interact with qubits, adding complexity to multi-qumode interactions. While a SWAP operation in DV systems typically requires 3 CNOT gates, in CV systems, it begins with a Beam-Splitter gate \( BS(\pi, 0) \) with parameters $\theta = \pi$ and $\varphi = 0$:

\begin{align}
BS(\pi, 0)\ket{\Phi_a, \Phi_b} = e^{-i\frac{\pi}{2}(a^\dagger a + b^\dagger b)} \ket{\Phi_b, \Phi_a}
\label{eq:bs}
\end{align}

To cancel the phase change from the Beam-Splitter (BS) gate, two Phase Space rotation gates, \( e^{-i\frac{\pi}{2}\hat{n}_a} \) and \( e^{-i\frac{\pi}{2}\hat{n}_b} \), are added. Thus, a \textbf{qumode SWAP gate} consists of \textbf{one Beam-Splitter (BS) gate} and \textbf{two Phase Space rotation gates}, with the latter having very low latency compared with a BS gate. This qumode SWAP primitive enables qumode movement for interactions with other qumodes or qubits. For qubit interactions involving qumodes, the qumodes can be moved to qubits. Multi-qubit interactions require primitives like Controlled Parity gates, discussed in Section~\ref{sec:multi-qubit-gate-synthesis}.

In physical implementations, connectivity constraints between qumodes, qubits, and qumode-qubit pairs can be summarized as the following three mapping challenges:

\begin{itemize}
    \item \textbf{Qumode-qumode Mapping. }  
    Interactions are limited to adjacent qumodes. For non-adjacent qumodes, qumode SWAP gates are used, with routing optimized to minimize such SWAPs.  
    \item \textbf{Qubit-qumode Mapping. }
    Each qumode interacts only with its associated qubit. For interactions with other qumodes, adjacency is established by moving qumodes, similar to qumode-qumode mapping. 
    \item \textbf{Qubit-qubit Mapping. }
    Qubits interact indirectly via an ancilla qumode, which is moved between qubits to mediate interactions and complete gate operations. 
\end{itemize}
 
\subsubsection{Optimized Ancilla Qumode Routing for Qubit-qubit Interactions}
\label{sec:TSPformulation}
As direct qubit-qubit interactions are limited by hardware constraints, we use ancilla qumodes to mediate interactions via phase kickback (Equation~\ref{eq:multiqubitgate}). The main challenge is to find the ancilla qumode and optimize its path to interact with target qubits efficiently. This \textit{Optimized Ancilla Qumode Routing Problem} seeks to minimize path costs, measured by qumode SWAP operations, by finding the shortest path for a specific ancilla qumode to visit all qubits in \( S \subseteq V \) on an undirected graph \( G = (V, E) \).

Fig.~\ref{fig:ancilla_routing} compares an arbitrary alphabetical routing path with an optimized ancilla qumode routing path, highlighting its significant impact on the overall SWAP cost:

\begin{figure}[htbp]
    \centering
    \includegraphics[width=0.3\textwidth]{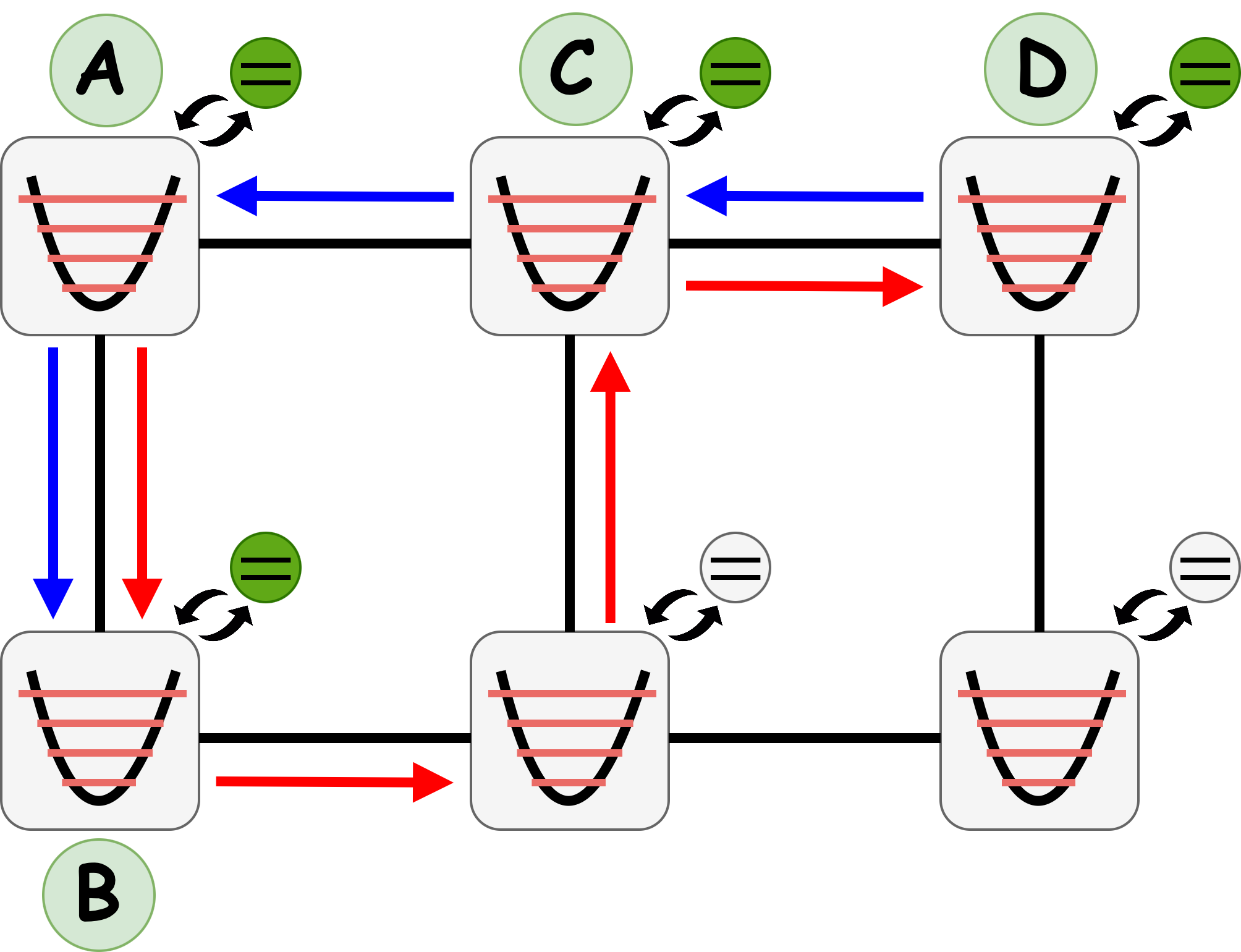}
    
    \caption{
    In the Optimized Ancilla Qumode Routing example, the red path represents alphabetical routing (\(A \rightarrow B \rightarrow C \rightarrow D\)) with a SWAP cost of 4. The blue path shows optimized routing (\(D \rightarrow C \rightarrow A \rightarrow B\)), using the Qumode at \(D\) as the ancilla, reducing the SWAP cost to 3.}
\Description{
This figure illustrates an Optimized Ancilla Qumode Routing example. There are four nodes labeled A, B, C, and D, each representing a Qumode. The red path follows the alphabetical routing sequence A to B to C to D, with a SWAP cost of 4. The blue path represents an optimized routing sequence D to C to A to B, which uses the Qumode at D as the ancilla, reducing the SWAP cost to 3. Arrows indicate the direction of the routing paths, with red arrows showing the alphabetical route and blue arrows showing the optimized route.
}

    \label{fig:ancilla_routing}
\end{figure}

The \textit{Optimized Ancilla Qumode Routing Problem} can be reformulated as a relaxed \textit{Hamiltonian Path Problem}, similar to a modified \textit{Traveling Salesman Problem} (TSP). Unlike the closed-path TSP, this problem allows revisiting vertices and does not require returning to the starting vertex.

We construct a complete graph \( G' = (V, E') \) with the same vertex set \( V \) as the original graph \( G \). Each edge in \( E' \) is weighted by \( w_{ij} \), the shortest path cost between \( v_i \) and \( v_j \) in \( G \), mapped to the edge sequence in \( G \). As shortest paths in \( G \) are polynomial-time computable, constructing \( G' \) is also polynomial in complexity.

Given the NP-hardness of the Hamiltonian Path Problem, we propose a multi-level solution strategy for scalability, as heuristic solutions can be more practical for larger problem instances.

We use \textit{Christofides Algorithm}, which constructs a minimum spanning tree, finds a minimum-weight perfect matching among the odd-degree vertices, and combines them to form an Eulerian circuit, which can guarantee a near-optimal solution with polynomial time complexity, making it suitable for large-scale qumode routing scenarios. However, we also explore the other Heuristic-based algorithms, such as \textit{Threshold Accepting Algorithm}, which can perform better in large-scale Pauli Strings compilation. 

Also, for larger-scale problems, \textit{Heuristic Algorithms} can provide \textbf{efficient} solutions, also making them well-suited for \textbf{specific physical circuits optimization goal} in Hybrid CV-DV systems to provide high extensibility and scalability.

\textbf{Dynamic Qubit Floating:} While the only native SWAP gates are for qumodes when addressing the connectivity constraints.  Qubit movement can also be considered. As discussed by Liu \etal \cite{liu2024hybridoscillatorqubitquantumprocessors}, CNOT gates can be implemented in the CV-DV architecture where qubits are not directly connected. We can essentially implement a qubit-qubit SWAP using 3 CNOT gates. However, it would require the usage of ancilla qumode, as the qubits are indirectly interacting via the ancilla qumodes. However, since qubit-qubit SWAP is not native, it takes 12 control displacement gates and 12 qumode-SWAP gates. It is almost 24 times the cost of the beamsplitter (qumode-SWAP) gate. It could only be useful when there are frequent interactions on a set of qubits, while the set of qubits is scattered very far from each other. This depends on the active-qubit pattern in the Pauli-string terms and the qubit-controlled qumode gates. Although it may not be useful in some applications, we still integrate this design into our compiler. We calculate the average shortest distance for active qubits within a Pauli-string, and if it exceeds a certain threshold, we trigger a clustering procedure and move qubits into a connected component.  It offers more flexibility for specific long-range interactions. This is as if qubits can float around rather than being at their fixed locations. We name this approach as \textbf{floating qubit}.  

Given hardware constraints and optimization flexibility, using an ancilla qumode or using qubit floating for multi-qubit gate synthesis offers three key advantages: \textbf{Resource Flexibility:} Any available qumode can serve as the ancilla without requiring a specific initial state, enabling efficient resource utilization. \textbf{Operational Efficiency:} The ancilla qumode can be reused across multiple gates, with Controlled Parity gates applied in any order before and after Displacement gates.
\textbf{Optimization Potential:} Ancilla selection, Ancilla/Qubit routing, and task assignment can be optimized for resource reuse and circuit partitioning.

\subsubsection{Mapping and Routing for General Case}
\label{sec:maproute}

Now we handle the case when the CVDV-QASM representation contains both Pauli strings and other CV-DV basis gates. Pauli-strings are processed separately. The general idea is to keep a working frontier -- the set of gates (Pauli-string gates) whose dependence has been resolved, as well as a scheduled\_worklist data structure containing gates that have already been scheduled so far. Then we repeat the following.  
\begin{enumerate}
    \item Execute any gate, including Pauli-string gates that are executable from the frontier. We prioritize the scheduling of Pauli-string gates because, by Trotterization, each Pauli-string gate (block) can run in any order. We use different ranking functions to choose the set of Pauli-strings to schedule next, and the ranking function either prioritizes the non-identity qubit number or the time stamp of the latest finished gate in that set of active qubits, to improve the parallelism. Then for the rest of the CV-DV basis gates, they are executable if their involved qubits/qumodes are connected. Once we schedule a sequence of gates, we add them to the scheduled\_worklist.and update the frontier.  
    \item For the rest of the gates in the frontier, pick SWAPs that will help at least one gate in the frontier. Rank all the possible SWAPs. Then select one SWAP and add it to the scheduled\_worklist. 
    \item Go back to Step 1 or terminate if there is no gate in the work frontier. 
\end{enumerate}

Once the program completes, the scheduled\_worklist implies the order of the compiled physical gates. For the cost function of ranking candidate SWAP gates, we use a method similar to the Sabre qubit mapper \cite{li+:asplos19}. For the cost function of ranking Pauli-strings, as mentioned above, we use either the minimum-active-qubit-number metric or the minimum-total-depth metric (sum up the depths of all active qubits in that Pauli-string) to parallelize as much as possible.

 We also design a specific coupling map data structure for hybrid CV-DV systems. Currently, there are three versions of hybrid CV-DV quantum processors: superconducting, trapped ion, and neutral atom. We do not consider the neutral atom architecture as it does not allow connectivity among qubits. Our coupling map can support both superconducting and trapped ion CV-DV architecture. 

\section{Evaluation}
\label{sec:eval}

\subsection{Experimental Setup}
\label{sec:expsetup}

Our evaluation focuses on two key aspects: {Pauli-string Hamiltonian (qubit-only) in a hybrid CV-DV architecture} and {general compilation support for Hamiltonian simulation}. First, we analyze the performance of compiling qubit-only Pauli-string Hamiltonians within the hybrid CV-DV architecture, evaluating our solutions for generating multi-qubit interactions via ancilla qumode(s). Second, we demonstrate, for the first time, the complete synthesis and compilation of given Hamiltonian models that may consist of both Fermions and Bosons, or just Bosons. This is for evaluating both the gate synthesis component and the hardware component. 

\textbf{Metrics.}  
We employ three primary metrics to assess the compiler's performance: \textbf{1-op Gate}, \textbf{2-op Gate}, and \textbf{Depth}. The \textbf{1-op Gate} and \textbf{2-op Gate} metrics count the number of single-operand and two-operand gates, respectively, reflecting interactions such as qubit-qumode or qumode-qumode operations. The \textbf{Depth} metric measures the number of layers of quantum gates after compilation, representing the complexity of the compiled circuit. Using data from Table~\ref{tab:gateSetGrouped}, we approximate the cost of a 2-op gate as equivalent to 20x that of a 1-op Gate and set 1-op gate latency to 1 when evaluating depth. For general Hamiltonian simulation compilation, the metric depth is for the final physical circuits that satisfy connectivity constraints rather than for the logical circuit. 

\textbf{Benchmarks.}
We select benchmarks of various sizes and applications from practical, real-world Hamiltonian models. For the fermion-only Hamiltonian, using the PySCF~\cite{sun2007python} software package, we constructed the Hamiltonians for seven distinct molecules: LiH, BeH$_2$, ethylene, NH$_3$, C$_2$, N$_2$, and H$_2$O, under minimal basis STO-3G.

We also consider different spin-orbital-to-electron ratios, resulting in 20 benchmarks that comprise a total of 600 to 1900 Pauli strings. For example, H$_2$O(8,12) and H$_2$O(10,14) represent water molecules with 8 electrons and 12 spin orbitals, and 10 electrons and 14 spin orbitals, respectively. The size of the molecular system affects the overall complexity of the simulation, and different encoding schemes can lead to variations in the number of required operations and the resulting circuit depth.

 We also include six general models, such as the Kerr nonlinear oscillator Hamiltonian, the $\mathbb{Z}_2$-Higgs model, the Bose-Hubbard model, the Hubbard-Holstein model, the Heisenberg model, and the electronic-vibration coupling Hamiltonians. They could include interactions of boson matters or fermion-boson matters.

    For each of the Hamiltonians, the fermion creation/annihilation operators must first be converted into qubit operators, in order to convert them into the form in Equation \ref{eq:HamiltonianRepresentation} before we perform further synthesis and scheduling. We employ both Jordan-Wigner~\cite{jordan1993paulische} (JW) and Bravyi-Kitaev~\cite{bravyi2002fermionic} (BK) encoding schemes to map fermionic operators to qubit operators~\cite{tilly+:arxiv22}. JW is highly regular, but its operator length grows linearly with the system size $N$($O(N)$). In contrast, the Bravyi-Kitaev encoding employs a more intricate tree-based structure to store parity information, thereby reducing the Pauli-weight, but with less regularity.

The Kerr nonlinear oscillator Hamiltonian ($H_1$) is a single-qumode Hamiltonian that includes a Kerr nonlinear term, adding complexity to its decomposition and compilation:

\begin{equation}
    H_1 = \omega a^{\dagger} a + \frac{\kappa}{2} (a^{\dagger})^2 a^2.
\end{equation}

The $\mathbb{Z}_2$-Higgs model ($H_2$) is a hybrid Hamiltonian featuring multi-mode interactions, including qumode-qumode and qubit-qubit interactions, making it representative of hybrid CV-DV systems:

\begin{equation}
    H_2 = - g \sum_{i=1}^{L-1} {X}_{i,i+1} + U \sum_{i=1}^{L} \hat{n}_i^2 - J \sum_{i=1}^{L-1}\left( \hat{a}^\dagger_{i} {Z}_{i,i+1} \hat{a}_{i+1} + \mathrm{h.c.} \right),
\end{equation}
where $\hat{n}_i = a_i^\dagger a_i$ is the bosonic number operator, and ${X}_{i,i+1}$ and ${Z}_{i,i+1}$ are Pauli X and Z operators linking sites $i$ and $i+1$, respectively, each can be represented using one qubit operator. The term ``h.c.'' represents the hermitian conjugate of the term right before the addition sign. 

The Bose-Hubbard model ($H_3$) is a lattice model whose structure varies with different lattice configurations, making it suitable for higher-dimensional physical systems. Also, it is a pure Hamiltonian with only qumode-qumode interaction (may need ancilla qubits for higher order annihilation and creation operators), making it representative of pure qumode Hamiltonian systems:

\begin{equation}
    H_3 = -t \sum_{i, j } (b_i^\dagger b_j + b_j^\dagger b_i) + \frac{U}{2} \sum_i \hat{n}_i (\hat{n}_i - 1) - \mu \sum_i \hat{n}_i,
\end{equation}
where $\hat{n}_i = b_i^\dagger b_i$, and $b_i^\dagger$ and $b_i$ are creation and annihilation operators.

The Hubbard-Holstein model ($H_4$) involves both fermionic and bosonic operators, adding complexity due to multiple mappings of fermionic operators to qubit Pauli operators. Here, $b^\dagger_{i}$ ($b_i$) denotes the bosonic creation (annihilation) operator at site $i$, while $c_{i,\sigma}^\dagger$ ($c_{i,\sigma}$) represents the fermionic counterpart. The number operator $\hat{n}_{i,\sigma} = c_{i,\sigma}^\dagger c_{i,\sigma}$ counts fermions with spin $\sigma = \uparrow, \downarrow$ at site $i$. We applied Jordan-Wigner encoding, but, due to space constraints, do not expand it into Pauli and Bosonic annihilation operator form:

\begin{equation}
    H_4 = -t \sum_{ i,j , \sigma} c_{i,\sigma}^\dagger c_{j,\sigma} + U \sum_i \hat{n}_{i,\uparrow} \hat{n}_{i,\downarrow} + \sum_i b_i^\dagger b_i + g \sum_{i,\sigma} \hat{n}_{i,\sigma} (b_i^\dagger + b_i),
\end{equation}

The Hamiltonian $H_5$ describes the interaction between discrete electronic states (qubits) and vibrational modes (qumodes), capturing complex electron-phonon coupling effects in molecular systems. Such models are essential for simulating quantum dynamics in one-dimensional (1D) chromophore arrays, where electronic excitations are strongly coupled to vibrational environments. Applications span light-harvesting complexes, organic semiconductors, and photosynthetic systems~\cite{vu2025computational}. The full Hamiltonian is decomposed as:

\begin{equation}
    H_5 = \sum_{\gamma = 1}^N \left[ H_0^{(\gamma)} + H_1^{(\gamma)} + H_2^{(\gamma)} \right]
    \label{eq:H5-total}
\end{equation}

Here, $H_0^{(\gamma)}$ represents local non-interacting terms, $H_1^{(\gamma)}$ captures dispersive coupling between qubits and modes, and $H_2^{(\gamma)}$ includes inter-chromophore interactions and vibrationally modulated couplings. Each component is defined below:

\begin{equation}
    H_0^{(\gamma)} = 
        \omega_{\gamma_0} b^\dagger_{\gamma_0} b_{\gamma_0} 
      + \omega_{\gamma_1} b^\dagger_{\gamma_1} b_{\gamma_1}
      - \frac{\omega_{q\gamma_0}}{2} \sigma^z_{\gamma_0} 
\end{equation}

\begin{equation}
          H_1^{(\gamma)} = 
        -\frac{\chi_{\gamma_0}}{2} b^\dagger_{\gamma_0} b_{\gamma_0} \sigma^z_{\gamma_0}
        + \frac{g_{cd,\gamma_0}}{2}(b_{\gamma_0} + b^\dagger_{\gamma_0}) \sigma^z_{\gamma_0} 
\end{equation}

\begin{align}
    H_2^{(\gamma)} &= 
        \frac{g_{cd,\gamma_1}}{2}(b_{\gamma_1} + b^\dagger_{\gamma_1}) \sigma^z_{\gamma_0} \notag \\
        &\quad + \frac{g_{\gamma_0,(\gamma\!-\!1)_0}}{4}
        \left( \sigma^x_{\gamma_0} \sigma^x_{(\gamma\!-\!1)_0} 
             + \sigma^y_{\gamma_0} \sigma^y_{(\gamma\!-\!1)_0} \right) \notag \\
        &\quad + \frac{g_{\gamma_0,(\gamma\!+\!1)_0}}{4}
        \left( \sigma^x_{\gamma_0} \sigma^x_{(\gamma\!+\!1)_0} 
             + \sigma^y_{\gamma_0} \sigma^y_{(\gamma\!+\!1)_0} \right) \notag \\
        &\quad + \frac{g_{\gamma_0,(\gamma\!-\!1)_0,\gamma_1}}{4}
        \left( \sigma^x_{\gamma_0} \sigma^x_{(\gamma\!-\!1)_0} 
             + \sigma^y_{\gamma_0} \sigma^y_{(\gamma\!-\!1)_0} \right)(b_{\gamma_1} + b^\dagger_{\gamma_1}) \notag \\
        &\quad + \frac{g_{\gamma_0,(\gamma\!+\!1)_0,\gamma_1}}{4}
        \left( \sigma^x_{\gamma_0} \sigma^x_{(\gamma\!+\!1)_0} 
             + \sigma^y_{\gamma_0} \sigma^y_{(\gamma\!+\!1)_0} \right)(b_{\gamma_1} + b^\dagger_{\gamma_1})
\end{align}

\begin{table*}[htbp]
\renewcommand{\arraystretch}{1.2}
\resizebox{1\textwidth}{!}{

\begin{tabular}{|c|c|c|ccc|ccc|ccc|}
\hline
\multirow{2}{*}{\textbf{Molecules}} & \multirow{2}{*}{Mapping} & \multirow{2}{*}{\# Pauli Strings} & \multicolumn{3}{c|}{Christofides Algorithm(Qumode SWAP)}                  & \multicolumn{3}{c|}{Threshold(Qumode SWAP)}          & \multicolumn{3}{c|}{Threshold(Floating Qubit)}         

\\ \cline{4-12} 
      &                          &    & \multicolumn{1}{c|}{1-op Gate} & \multicolumn{1}{c|}{2-op Gate} & Duration   & \multicolumn{1}{c|}{1-op Gate} & \multicolumn{1}{c|}{2-op Gate} & Duration            & \multicolumn{1}{c|}{1-op Gate} & \multicolumn{1}{c|}{2-op Gate} & Duration              
      
      \\ \hline
\multirow{2}{*}{LiH (4, 12)}        & BK                       & 631& \multicolumn{1}{c|}{41472}     & \multicolumn{1}{c|}{30920}     & 546597  & \multicolumn{1}{c|}{39192}     & \multicolumn{1}{c|}{29780}     & 518510 (5.14\%)  & \multicolumn{1}{c|}{39360}     & \multicolumn{1}{c|}{29864}     & 520580 (4.76\%)    

\\ \cline{2-12} 
      & JW                       & 631& \multicolumn{1}{c|}{36344}     & \multicolumn{1}{c|}{27384}     & 458441  & \multicolumn{1}{c|}{33888}     & \multicolumn{1}{c|}{26156}     & 432246 (5.71\%)  & \multicolumn{1}{c|}{34352}     & \multicolumn{1}{c|}{26472}     & 444074 (3.13\%)   
      
      \\ \hline
\multirow{2}{*}{BeH2 (6, 14)}       & BK                       & 666& \multicolumn{1}{c|}{49520}     & \multicolumn{1}{c|}{35608}     & 628278  & \multicolumn{1}{c|}{46576}     & \multicolumn{1}{c|}{34136}     & 587180 (6.54\%)  & \multicolumn{1}{c|}{48210}     & \multicolumn{1}{c|}{35184}     & 644966 (-2.66\%)   

\\ \cline{2-12} 
      & JW                       & 666& \multicolumn{1}{c|}{44428}     & \multicolumn{1}{c|}{33652}     & 562822  & \multicolumn{1}{c|}{42092}     & \multicolumn{1}{c|}{32484}     & 535692 (4.82\%)  & \multicolumn{1}{c|}{43384}     & \multicolumn{1}{c|}{33508}     & 585151 (-3.97\%)   
      
      \\ \hline
\multirow{2}{*}{Ethylene (10, 16)}  & BK                       & 789& \multicolumn{1}{c|}{54976}     & \multicolumn{1}{c|}{40688}     & 678566  & \multicolumn{1}{c|}{52368}     & \multicolumn{1}{c|}{39384}     & 650664 (4.11\%)  & \multicolumn{1}{c|}{57602}     & \multicolumn{1}{c|}{43408}     & 769017 (-13.33\%)  

\\ \cline{2-12} 
      & JW                       & 789& \multicolumn{1}{c|}{57312}     & \multicolumn{1}{c|}{43496}     & 712622  & \multicolumn{1}{c|}{54304}     & \multicolumn{1}{c|}{41992}     & 681511 (4.37\%)  & \multicolumn{1}{c|}{58066}     & \multicolumn{1}{c|}{44608}     & 796156 (-11.72\%)  
      
\\ \hline
\multirow{2}{*}{NH3 (8, 12)}        & BK                       & 915& \multicolumn{1}{c|}{61564}     & \multicolumn{1}{c|}{45592}     & 810401  & \multicolumn{1}{c|}{58164}     & \multicolumn{1}{c|}{43892}     & 773912 (4.5\%)   & \multicolumn{1}{c|}{58268}     & \multicolumn{1}{c|}{43944}     & 769401 (5.06\%)    

\\ \cline{2-12} 
      & JW                       & 915& \multicolumn{1}{c|}{53680}     & \multicolumn{1}{c|}{39960}     & 656519  & \multicolumn{1}{c|}{50336}     & \multicolumn{1}{c|}{38288}     & 615952 (6.18\%)  & \multicolumn{1}{c|}{53002}     & \multicolumn{1}{c|}{39684}     & 699905 (-6.61\%)  
      
      \\ \hline
\multirow{2}{*}{C2 (10, 16)}        & BK                       & 1177                             & \multicolumn{1}{c|}{87200}     & \multicolumn{1}{c|}{63768}     & 1057647 & \multicolumn{1}{c|}{82584}     & \multicolumn{1}{c|}{61460}     & 1011085 (4.4\%)  & \multicolumn{1}{c|}{85998}     & \multicolumn{1}{c|}{64196}     & 1111745 (-5.11\%)  

\\ \cline{2-12} 
      & JW                       & 1177                             & \multicolumn{1}{c|}{88960}     & \multicolumn{1}{c|}{66968}     & 1100289 & \multicolumn{1}{c|}{83832}     & \multicolumn{1}{c|}{64404}     & 1057411 (3.9\%)  & \multicolumn{1}{c|}{87568}     & \multicolumn{1}{c|}{67028}     & 1190209 (-8.17\%) 
      
      \\ \hline
\multirow{2}{*}{N2 (10, 16)}        & BK                       & 1177                             & \multicolumn{1}{c|}{88792}     & \multicolumn{1}{c|}{65004}     & 1074193 & \multicolumn{1}{c|}{84488}     & \multicolumn{1}{c|}{62852}     & 1039060 (3.27\%) & \multicolumn{1}{c|}{87078}     & \multicolumn{1}{c|}{65176}     & 1150102 (-7.07\%) 

\\ \cline{2-12} 
      & JW                       & 1177                             & \multicolumn{1}{c|}{88752}     & \multicolumn{1}{c|}{67056}     & 1106544 & \multicolumn{1}{c|}{84232}     & \multicolumn{1}{c|}{64796}     & 1063242 (3.91\%) & \multicolumn{1}{c|}{89512}     & \multicolumn{1}{c|}{68528}     & 1216367 (-9.92\%)  
      
      \\ \hline
\multirow{2}{*}{H2O (8, 12)}        & BK                       & 1219                             & \multicolumn{1}{c|}{80004}     & \multicolumn{1}{c|}{59516}     & 1048788 & \multicolumn{1}{c|}{75860}     & \multicolumn{1}{c|}{57444}     & 1001186 (4.54\%) & \multicolumn{1}{c|}{75940}     & \multicolumn{1}{c|}{57484}     & 997611 (4.88\%)    

\\ \cline{2-12} 
      & JW                       & 1219                             & \multicolumn{1}{c|}{70200}     & \multicolumn{1}{c|}{52696}     & 862689  & \multicolumn{1}{c|}{66336}     & \multicolumn{1}{c|}{50764}     & 817900 (5.19\%)  & \multicolumn{1}{c|}{70540}     & \multicolumn{1}{c|}{52908}     & 937287 (-8.65\%)  
      
      \\ \hline
\multirow{2}{*}{H2O (10, 14)}       & BK                       & 1654                             & \multicolumn{1}{c|}{122864}    & \multicolumn{1}{c|}{88204}     & 1543297 & \multicolumn{1}{c|}{116576}    & \multicolumn{1}{c|}{85060}     & 1469419 (4.79\%) & \multicolumn{1}{c|}{122572}    & \multicolumn{1}{c|}{91040}     & 1620031 (-4.97\%) 

\\ \cline{2-12} 
      & JW                       & 1654                             & \multicolumn{1}{c|}{111332}    & \multicolumn{1}{c|}{83068}     & 1377143 & \multicolumn{1}{c|}{105180}    & \multicolumn{1}{c|}{79992}     & 1319874 (4.16\%) & \multicolumn{1}{c|}{113576}    & \multicolumn{1}{c|}{84904}     & 1553009 (-12.77\%) 
      
      \\ \hline
\multirow{2}{*}{NH3 (8, 14)}        & BK                       & 1734                             & \multicolumn{1}{c|}{131716}    & \multicolumn{1}{c|}{94452}     & 1657461 & \multicolumn{1}{c|}{124404}    & \multicolumn{1}{c|}{90796}     & 1547533 (6.63\%) & \multicolumn{1}{c|}{129320}    & \multicolumn{1}{c|}{93968}     & 1700487 (-2.6\%)  

\\ \cline{2-12} 
      & JW                       & 1734                             & \multicolumn{1}{c|}{118260}    & \multicolumn{1}{c|}{88428}     & 1484725 & \multicolumn{1}{c|}{111620}    & \multicolumn{1}{c|}{85108}     & 1395552 (6.01\%) & \multicolumn{1}{c|}{120348}    & \multicolumn{1}{c|}{90228}     & 1622020 (-9.25\%) 
      
      \\ \hline
\multirow{2}{*}{C2 (12, 18)}        & BK                       & 1884                             & \multicolumn{1}{c|}{175612}    & \multicolumn{1}{c|}{126176}    & 2212185 & \multicolumn{1}{c|}{167756}    & \multicolumn{1}{c|}{122248}    & 2141333 (3.2\%)  & \multicolumn{1}{c|}{171994}    & \multicolumn{1}{c|}{131024}    & 2381938 (-7.67\%) 

\\ \cline{2-12} 
      & JW                       & 1884                             & \multicolumn{1}{c|}{155140}    & \multicolumn{1}{c|}{117292}    & 1937459 & \multicolumn{1}{c|}{147340}    & \multicolumn{1}{c|}{113392}    & 1857982 (4.1\%)  & \multicolumn{1}{c|}{163376}    & \multicolumn{1}{c|}{127668}    & 2361642 (-21.89\%) 
      
\\ \hline
\end{tabular}

}
\caption{Compilation latency results for multi-Pauli exponentials. Gate latencies are assigned according to Table~\ref{tab:gateSetGrouped}, where single-qubit (1-op) gates have a latency of 1 unit (20 nanoseconds) and two-qubit (2-op) gates incur a latency of 20 units. The Christofides algorithm is used as the performance baseline. Christofides-based routing is used as the baseline. Duration results show absolute latency and percentage reduction relative to the baseline.}
\label{tab:table1}
\end{table*}

The Heisenberg model ($H_6$) describes qubit-qubit interactions in a spin chain or lattice system. Here, $H_6$ represents a chain of $N$ qubits coupled via exchange interactions in the $x, y, z$ directions. The absence of qubit-qumode and qumode-qumode interactions simplifies the model, making it ideal for understanding pure qubit dynamics. The Heisenberg Hamiltonian has broad applications in condensed matter physics, quantum magnetism, and quantum computing, serving as a foundation for entanglement, spin transport, and quantum phase transitions, which is given by:
\begin{align}
    H_6 = -\frac{1}{2} \sum_{j=1}^N(J_x X_j X_{j+1} + J_y Y_j Y_{j+1} + J_z Z_j Z_{j+1} + h Z_j)
\end{align}

These Hamiltonians exemplify hybrid CV-DV systems and present significant compilation challenges. For the first time, we have fully decomposed, synthesized, and compiled their physical circuits for hybrid CV-DV quantum computers.

\textbf{Baselines and Hardware Coupling Maps.}  
For multi-qubit interactions on a hybrid CV-DV architecture, since there is a prior compiler study, we compare different versions of our implementation, including the Christofides algorithm and the Threshold Accepting Heuristic-based approach, to analyze their effectiveness in handling such Hamiltonian Simulation scenarios. Additionally, we investigate the use of Floating Qubits within the hybrid CV-DV architecture, exploring innovative compilation strategies under this framework. Finally, we assess the scalability and extensibility of our methods, offering valuable insights for future research and development. For the coupling map, we adopt a lattice structure for qumode connectivity as shown in Fig. \ref{fig:hybridcvdvarch}, each qumode is connected to one qubit. Qubits are not connected. 

\subsection{Pauli String Synthesis using Ancilla Qumode and Floating Qubit}
\label{sec:evalpauli}

For Pauli string compilation, hybrid CV-DV systems introduce challenges in handling qubit-qubit interactions where direct qubit compilation is unavailable. This necessitates adaptations in our framework. One approach mediates multi-qubit interactions through qumodes, addressing the Optimized Ancilla Qumode Routing Problem. We explored two routing strategies: the Christofides Algorithm and the Threshold Accepting Algorithm, both efficient heuristics for the common traveling salesman problem (TSP). Additionally, we investigated the Floating Qubit method as an alternative architectural solution for hybrid systems. Detailed results are in Table~\ref{tab:table1}.  
Our benchmarks involve complex and representative Hamiltonians. While the Christofides Algorithm provides an efficient polynomial-time approximation, results indicate that the Threshold Accepting Algorithm achieves 3-7\% better optimization, reducing circuit depth by an average of 4.8\%. This demonstrates the effectiveness of heuristics in addressing complex compilation problems. Further improvements might be possible by incorporating additional heuristic objectives, such as architecture-specific optimizations.

We also evaluated the Floating Qubit method but found it largely ineffective across our benchmark suite, achieving optimal results in only two cases. In most benchmarks, it increased the average circuit depth by 6\%. This is attributed to the significant cost of Floating Qubits, which can reach up to 24x, making the depth minimization goal more challenging. However, flexible Floating Qubit strategies may still be useful in scenarios requiring high adaptability, such as resource reuse or circuit partitioning.

Table \ref{tab:table1} shows that JW outperforms BK in physical mapping and routing. It is because JW's Pauli strings have more regularity from block to block and hence better locality, thereby reducing the number of additional routing gates. Therefore, we use JW encoding for fermion operators in all subsequent experiments.  

\subsection{General Hamiltonian Simulation Compilation}
\subsubsection{End-to-end Compilation}
For the evaluation of general Hamiltonian simulation, we compiled a diverse set of six Hamiltonian models. These models are not only representative in hybrid CV-DV systems but also pose significant complexity during compilation, making them ideal for evaluating our general-purpose compiler. We further tested performance across varying lattice sizes to assess scalability. The detailed results are shown in Table~\ref{tab:table2}.

    Table~\ref{tab:table2} shows the results at different stages of compilation. The number of Pauli strings and hybrid CV-DV gates is from the intermediate representation using our domain-specific language, \textbf{CVDV-QASM}. The ``Total Gate Count'' and ``Duration'' are from the final compiled circuits after decomposition, mapping, and routing.

These representative Hamiltonian models highlight the compiler's capability to support practical quantum computing applications, offering a valuable framework for benchmarking and reference in future research. Key considerations included successful decomposition, gate count, and acceptable depth metrics. Based on our Rule-Based Recursive Template Matching Mechanism, the framework can be further optimized for architecture-specific constraints or fidelity improvements, providing a foundation for future advancements in Hamiltonian simulation compilation.

\subsubsection{Hit Rate Analysis for Decomposition Rules}
\label{sec:hitrate}
We conducted a detailed decomposition analysis by measuring the ``hit rate'' of each rule, quantifying its relative importance during synthesis. Each unique term in a summation \(\Sigma\) was counted once per Hamiltonian model, regardless of index variations. For complex expressions (e.g., \(a^\dagger a^\dagger a a\), \(a^\dagger a a^\dagger a\)), which often require multiple recursive applications, we normalized rule usage per Hamiltonian model. This normalized frequency is defined as the rule's ``hit rate.''

Rule 16, which corresponds to native basis gate synthesis, appears most frequently and nearly all decomposition paths eventually terminate at a basis gate. It accounts for 68.30\% of rule applications along successful paths, and 70.88\% overall. To better assess the relative contribution of other rules, we exclude Rule 16 and report the normalized hit rates for Rules 1-15 in Table~\ref{tab:hitRate}.

Rules 1 and 2 (Trotter and BCH decompositions) are heavily used during intermediate steps, together contributing over 28.80\% of the successful hit rate. Rules 14 and 15, which address multi-Pauli exponentials and multi-qubit-controlled qumode displacements, also show high hit rates, highlighting the importance of qubit-qumode interactions in certain Hamiltonian model. Rules 11 and 12 frequently appear in the successful paths as well, enabling bosonic operators to be decomposed into fine-grained gate-synthesizable forms via block-encoding templates, jointly contributing 10.80\% to the successful hit rate.

\subsubsection{Compilation Time Analysis}
We evaluated the compilation time of our framework using JW-mapped Pauli strings, considering the scale of Pauli string compilation for each molecule. Results are shown in Table~\ref{tab:time}. Measured in seconds, the compilation time covers the entire process, from Pauli string decomposition to physical circuit synthesis. While it generally increases with the number of Pauli strings, C2(12, 18) shows a significantly higher time due to the increased number of electrons and spin-orbitals, leading to a more complex system than just longer Pauli strings.

\begin{table*}[htbp]
\renewcommand{\arraystretch}{1}
\resizebox{0.8\textwidth}{!}{
\begin{tabular}{|c|c|c|c|c|c|c|}
\hline
\textbf{Hamiltonians} & \# Qubits & \# Qumodes & \# Pauli Strings (in IR) & \# Gates (in IR) & Final Gate Count & Duration \\ \hline
Kerr Nonlinear Hamiltonian & 19 (ancilla) & 1 & 0 & 409 & 595 & 3032 \\ \hline
\multirow{3}{*}{$Z_2$-Higgs Model} & 20 & 20 & 19 & 419 & 696 & 943 \\ \cline{2-7} 
 & 40 & 40 & 39 & 839 & 1472 & 1797 \\ \cline{2-7} 
 & 60 & 60 & 59 & 1259 & 2263 & 2721 \\ \hline
\multirow{3}{*}{Bose-Hubbard Model} & 20 (ancilla) & 20 & 0 & 820 & 2935 & 15353 \\ \cline{2-7} 
 & 40 (ancilla) & 40 & 0 & 2440 & 15313 & 62548 \\ \cline{2-7} 
 & 60 (ancilla) & 60 & 0 & 4860 & 36705 & 155872 \\ \hline
Hubbard-Holstein Model & 40 & 20 & 3100 & 100 & 688011 & 4509157 \\ \hline
\multirow{3}{*}{Electronic-Vibration Coupling} & 20 & 40 & 96 & 1392 & 5635 & 28551 \\ \cline{2-7} 
 & 40 & 80 & 196 & 2852 & 12655 & 54952 \\ \cline{2-7} 
 & 60 & 120 & 296 & 4312 & 19684 & 81729 \\ \hline
\multirow{3}{*}{Heisenberg Model} & 20 & 0 & 77 & 0 & 2218 & 4122 \\ \cline{2-7} 
 & 40 & 0 & 157 & 0 & 4676 & 4398 \\ \cline{2-7} 
 & 60 & 0 & 237 & 0 & 7200 & 7074 \\ \hline
\end{tabular}
}
\caption{
        Compilation results for general Hamiltonian simulations. For each model, the table reports (a) the number of qubits and qumodes utilized, (b) the component counts in intermediate representation, i.e., the number of Pauli strings and hybrid CV-DV gates, using our domain-specific language, \textbf{CVDV-QASM}, and (c) the ``Total Gate Count'' and ``Duration'' of the final compiled circuit. Operation durations are estimated by assigning 1 time unit (20 nanoseconds) to each single-qubit or single-qumode gate and 20 time units to each hybrid CV-DV gate and multi-qumode gate, according to Table~\ref{tab:gateSetGrouped}.
}

\label{tab:table2}
\end{table*}

\begin{table}[htbp]
    \centering
    \resizebox{0.47\textwidth}{!}{
    \begin{tabular}{|c|c|c|c|c|c|c|c|c|}
        \hline
        {\textbf{Rule}} & {\textbf{Success}} & {\textbf{Total}} & {\textbf{Rule}} & {\textbf{Success}} & {\textbf{Total}} & {\textbf{Rule}} & {\textbf{Success}} & {\textbf{Total}}\\ \hline
        {No. 1}  & {9.50\%} & {3.83\%}  & {No. 6}  & {1.35\%} & {0.97\%}  & {No. 11} & {0.34\%} & {4.47\%}  \\ \hline
        {No. 2}  & {19.33\%} & {23.77\%}  & {No. 7}  & {1.35\%} & {0.57\%}  & {No. 12} & {5.40\%} & {12.78\%}  \\ \hline
        {No. 3}  & {0.00\%} & {0.00\%}  & {No. 8}  & {8.15\%} & {1.47\%}  & {No. 13} & {5.40\%} & {11.48\%} \\ \hline
        {No. 4}  & {8.15\%} & {2.94\%}  & {No. 9}  & {0.00\%} & {1.47\%}  & {No. 14} & {15.77\%} & {17.17\%} \\ \hline
        {No. 5}  & {8.15\%} & {1.47\%}  & {No. 10} & {1.35\%} & {0.45\%}  & {No. 15} & {15.77\%} & {17.17\%} \\ \hline
    \end{tabular}
    }
    \caption{
        Normalized hit rates for decomposition Rules 1-15. Each entry includes both the \textit{Success Hit Rate} (calculated over successfully decomposed paths) and the \textit{Total Hit Rate} (all search attempts, including failures).
    }
    \vspace*{-1\baselineskip}
    \label{tab:hitRate}
\end{table}

\begin{table}[htb]
    \centering
    \resizebox{0.47\textwidth}{!}{
    \begin{tabular}{|c|c|c|c|c|c|}
    \hline
    \textbf{Molecule} & \textbf{\# Pauli Strings} & \textbf{Time(s)} & \textbf{Molecule} & \textbf{\# Pauli Strings} & \textbf{Time(s)} \\ \hline
    LiH (4, 12)       & 631           & 20.39   & BeH2 (6, 14)      & 666           & 33.33   \\ \hline
    Ethylene (10, 16) & 789           & 67.21   & NH3 (8, 12)       & 915           & 59.95   \\ \hline
    C2 (10, 16)       & 1177          & 184.49  & N2 (10, 16)       & 1177          & 205.48  \\ \hline
    H2O (8, 12)       & 1219          & 100.33  & H2O (10, 14)      & 1654          & 379.83  \\ \hline
    NH3 (8, 14)       & 1734          & 485.46  & C2 (12, 18)       & 1884          & 1,152.00 \\ \hline
    \end{tabular}
    }
    \caption{Compilation Time Analysis}
    \vspace*{-2\baselineskip}
    \label{tab:time}
\end{table}

\section{Related Work}

Existing tools, including Bosonic Qiskit \cite{stavenger+:hpec22}, StrawberryField \cite{killoran2019strawberryfields}, Perceval \cite{zhou2024bosehedral}, and Bosehedral \cite{maring2023generalpurposesinglephotonbasedquantumcomputing}, offer preliminary support for programming, simulating, and composing circuits for either domain-specific applications such as Gaussian Boson Sampling (GBS) or general bosonic circuits. Perceval allows users to build linear optical circuits from a collection of pre-defined components. Strawberry Fields and Bosonic Qikist provide Python libraries involving basic CV-DV gates, as well as simulate the programs written with these basic libraries.  However, none of this provides support for the simulation of the hybrid CV-DV Hamiltonian.

While Hamiltonian simulation on DV systems has been extensively studied \cite{li+:asplos22, li+:isca21, jin2023tetris}, compilation support for hybrid CV-DV systems remains underexplored. The unique properties of Bosonic hardware mean that Fermion-Boson interactions have not been thoroughly investigated. This involves converting Hamiltonian simulation algorithms into basis gates for both qubits and qumodes, implementing multi-qubit operations, and optimizing qubit-qumode mapping and routing to address connectivity and error mitigation challenges. Our work bridges this gap.

\section{Conclusions}

Our work introduces Genesis, the first comprehensive compilation framework for Hamiltonian simulation on hybrid CV-DV quantum processors. By leveraging a two-stage compilation approach: (1) decomposing hybrid Hamiltonians into universal basis gates, and (2) mapping them to hardware-constrained circuits, we enable efficient simulation of complex physical systems. Our tool has successfully compiled important  Hamiltonians, including the Bose-Hubbard model, $\mathbb{Z}_2-$Higgs model, Hubbard-Holstein model, and electron-vibration coupling Hamiltonians critical in domains like quantum field theory, condensed matter physics, and quantum chemistry. We also provide a domain-specific language (DSL) design to support Hamiltonian simulation on a hybrid CV-DV architecture. 

\begin{acks} 
This work is supported in part by grants from the Rutgers Research Council, NSF, and DOE. In particular, Y.L., H.Z., and J.B. are supported by the U.S. Department of Energy (DOE), Office of Science, Office of Advanced Scientific Computing Research (ASCR), under Award Number DE-SC0025384. Z.C., J.L., M.G., H.C., and E.Z. are supported by the DOE Award DE-SC0025563, the NSF Award CCF-2129872, and Rutgers Research Council Grant. J.B., Y.L., and H.Z. are supported in part by NSF OSI-2410675 (with a subcontract to NC State University from Duke University). H.Z. is also supported in part by NSF grants PHY-2325080 and OMA-2120757 (with a subcontract to NC State University from the University of Maryland). Any opinions, findings, conclusions, or recommendations expressed in this material are those of the authors and do not necessarily reflect the views of our sponsors.
\end{acks}

\bibliographystyle{ACM-Reference-Format}
\bibliography{main}

\end{document}